\documentclass[12pt]{article}

\makeatletter
 \@addtoreset{equation}{section}
 
\makeatother
\makeatletter
\newcount\@tempcntc
\def\@citex[#1]#2{\if@filesw\immediate\write%
                  \@auxout{\string\citation{#2}}\fi
  \@tempcnta\z@\@tempcntb\m@ne\def\@citea{}\@cite{\@for\@citeb:=#2\do
    {\@ifundefined
       {b@\@citeb}{\@citeo\@tempcntb\m@ne\@citea%
                   \def\@citea{,}{\bf ?}\@warning
       {Citation `\@citeb' on page \thepage \space undefined}}%
    {\setbox\z@\hbox{\global\@tempcntc0\csname b@\@citeb%
                     \endcsname\relax}%
     \ifnum\@tempcntc=\z@ \@citeo\@tempcntb\m@ne
       \@citea\def\@citea{,}\hbox{\csname b@\@citeb\endcsname}%
     \else
      \advance\@tempcntb\@ne
      \ifnum\@tempcntb=\@tempcntc
      \else\advance\@tempcntb\m@ne\@citeo
      \@tempcnta\@tempcntc\@tempcntb\@tempcntc\fi\fi}}\@citeo}{#1}}
\def\@citeo{\ifnum\@tempcnta>\@tempcntb\else\@citea\def\@citea{,}%
  \ifnum\@tempcnta=\@tempcntb\the\@tempcnta\else
   {\advance\@tempcnta\@ne\ifnum\@tempcnta=\@tempcntb%
     \else \def\@citea{--}\fi
    \advance\@tempcnta\m@ne\the\@tempcnta\@citea\the\@tempcntb}\fi\fi}
\makeatother
\begin{document}
%
\vspace*{-1.0cm}
\begin{center}
{\Large\bf Effects of new non-linear couplings in relativistic 
effective field theory}
\\[2.0cm]
M. Del Estal, M. Centelles, X. Vi\~nas and S.K. Patra \\[2mm]
{\it Departament d'Estructura i Constituents de la Mat\`eria,
     Facultat de F\'{\i}sica,
\\
     Universitat de Barcelona,
     Diagonal {\sl 647}, E-{\sl 08028} Barcelona, Spain}
\end{center}
%
\vspace*{2.0cm}
\begin{abstract}
We extend the relativistic mean field theory model of Sugahara and
Toki (TM1) by adding new couplings suggested by modern effective field
theories. An improved set of parameters (TM1*) is developed with the
goal to test the ability of the models based on effective field theory
to describe the properties of finite nuclei and, at the same time, to
be consistent with the trends of Dirac--Brueckner--Hartree--Fock
calculations at densities away from the saturation region. We compare
our calculations with other relativistic nuclear force parameters for
various nuclear phenomena.
\end{abstract}

\mbox{}



\pagebreak
%
\section{Introduction}
In the recent decades relativistic quantum field theory has been very
successful for the description of the nuclear many-body problem
\cite{Ma89,Ho84,An83,Br84,Ha87}. The relativistic models take care
{\em ab initio} of many natural phenomena which are practically absent
or have to be included in an {\em ad hoc} manner in the
non-relativistic formalism. Specifically, the relativistic mean field
(RMF) treatment of quantum hadrodynamics (QHD) \cite{Se86,Se97} has
become a popular way to deal with the nuclear physics problems. The
original linear $\sigma-\omega$ model of Walecka \cite{Wa74} was
complemented with cubic and quartic non-linearities of the $\sigma$
meson \cite{Bo77} (non-linear $\sigma-\omega$ model) to improve the
incompressibility and the finite nuclei results. Since these models
were proposed to be renormalizable, the scalar self-interactions were
limited to a quartic polynomial and scalar-vector and vector-vector
interactions were not allowed. Very recently, however, inspired by
effective field theory (EFT), Furnstahl, Serot and Tang
\cite{Fu96,Fu97} abandoned the idea of renormalizability and extended
the RMF theory by allowing other non-linear scalar-vector and
vector-vector self-interactions in addition to tensor couplings
\cite{Se97,Fu96,Fu97,Se96}. 

The effective field theory contains all the non-renormalizable
couplings consistent with the underlying symmetries of QCD. Since one
has to deal with an effective Lagrangian with an infinite number of
terms it is imperative to develop a suitable expansion scheme. In the
nuclear many-body problem the scalar ($\Phi$) and vector ($W$) meson
fields are normally small as compared with the nucleon mass ($M$) and
they vary slowly in finite nuclei. This means that the ratios
$\Phi/M$, $W/M$, $|\mbox{\boldmath $\nabla$}\Phi|/M^2$ and
$|\mbox{\boldmath $\nabla$} W|/M^2$ are the useful expansion
parameters \cite{Se97,Fu96,Fu97,Se96}. The concept of naturalness
\cite{Se97,Fu97}, i.e., that all the coupling constants written in an
appropriate dimensionless form should be of the order of unity, is
used to avoid ambiguities in the expansion. Then one can estimate the
contributions coming from different terms by counting powers in the
expansion parameters and truncating the Lagrangian at a given level of
accuracy. For the truncation to be consistent, the coupling constants
should exhibit naturalness and none should be arbitrarily dropped out
to the given order without additional symmetry arguments.

If the EFT Lagrangian is truncated at fourth order one recovers the
standard non-linear $\sigma-\omega$ model plus some additional
couplings \cite{Se97,Fu97}, with thirteen free parameters. In Ref.\
\cite{Fu97} these parameters have been fitted (sets G1 and G2) to
reproduce twenty-nine finite nuclei observables (binding energies,
charge form factors and spin-orbit splittings of magic nuclei). The
fits display naturalness and the results are not dominated by the last
terms retained. This evidence confirms the utility of the EFT concepts
and of the naturalness assumption, and shows that truncating the
effective Lagrangian at the first lower orders is justified. The EFT
approach has also been helpful to elucidate the empirical success of
the usual non-linear $\sigma-\omega$ models that have less free
parameters. It has been shown that the mean field phenomenology of
bulk and single-particle nuclear observables does not constrain all of
the parameters of the EFT model unambiguosly. That is, the constants
in the EFT model are underdetermined by the observables included in
the fits and several parameter sets with low $\chi^2$ can be found
\cite{Se97,Fu97,Fu97b,Cl98,Fu00}. An analysis of the particular impact
of each one of the new couplings arising in EFT on the determination
of the saturation properties of nuclear matter and on the nuclear
surface properties has been carried out in Ref.\ \cite{Es99}. Recent
applications of the EFT-based models include studies of pion-nucleus
scattering \cite{Cl98}, the nuclear spin-orbit force \cite{Fu98} and
asymmetric nuclear matter at finite temperature \cite{Wa00}.

On a more microscopic level, it is well known that non-relativistic
Brueckner--Goldstone calculations based on realistic NN potentials are
not able to give the right saturation density and binding energy of
infinite nuclear matter at the same time (Coester line) \cite{Co70}.
To obtain relatively correct values an additional repulsive part has
to be added. This can be achieved by working in the relativistic
framework: the Dirac--Brueckner--Hartree--Fock (DBHF) theory
\cite{Ma89,Ho84,An83,Br84,Ha87,Am92} introduces an extra density
dependence that allows to fit the NN phase shifts and to approach the
empirical equilibrium point of nuclear matter. The large scalar and
time-like vector self-energies of the DBHF calculations show two
interesting features: a rather small effect of the two-body
correlations and a weak momentum dependence, at least for not very
high densities. This suggests fitting the nuclear matter DBHF
self-energies by a much simpler RMF approach. This strategy was
carried out in the past using the $\sigma-\omega$ model with scalar
self-interactions \cite{Gm91,rashdan97}, and also including a quartic
vector self-interaction \cite{Gm92a} (as proposed by Bodmer
\cite{Bo89}). The outcoming parameter sets did not properly reproduce
the properties of finite nuclei. The saturation properties were close
to those of DBHF and it is known that they are not accurate enough, in
spite of the significant improvement over the non-relativistic BHF
results. 

Refs.\ \cite{Gm91,rashdan97,Gm92a} showed that the success of the
usual RMF model with only scalar self-interactions for describing the
saturation point and the data for finite nuclei is not followed by a
proper description of the trends of the DBHF scalar and vector
self-energies. This is caused mainly by a too restrictive treatment of
the $\omega-$meson term: while in the standard RMF model the vector
potential increases linearly with density and gets stronger, in DBHF
it bends down with density (see Figure~1 later). Moreover, the scalar
potential overestimates the DBHF result at high density in order to
compensate for the strong repulsion in the vector channel. This is the
reason for providing the wrong sign in the coupling constant of the
$\Phi^4$ term in most of the successful RMF parameter sets.
Furthermore, the equation of state becomes much steeper and soon
separates from the DBHF tendency when the density grows (see
Figure~2). Adding a quartic vector self-interaction remarkably
improves the behaviour of the vector and scalar potentials, softens
the equation of state and brings about a positive sign for the
$\Phi^4$ coupling \cite{Gm92a,Bo89,Su94}. In particular, Sugahara and
Toki \cite{Su94} took into account the non-linear term ($W^4$) in the
$\omega$ vector field and fitted the free parameters to the data for
several nuclei. Even with inclusion of the $W^4$ term they could not
get with a single parametrization, and at the same time, a positive
coupling constant of the $\Phi^4$ term and a quality for nuclei along
the periodic table similar to that of NL1 \cite{Re86}. Thus they
constructed two parameter sets, TM1 and TM2, both with a positive
$\Phi^4$ coupling constant. The TM2 set was designed for charge
numbers $Z \leq 20$ and the TM1 set for larger $Z$. TM1 was also
applied to calculate the equation of state and the structure of
neutron stars and supernovae \cite{Su94,Su95}. Apart from giving good
results for finite nuclei of $Z\geq 20$, TM1 agreed with the trends of
the nuclear matter DBHF calculations much better than the conventional
non-linear $\sigma-\omega$ sets (such as NL1 \cite{Re86} or NL3
\cite{Sh93}) owing to the vector self-interaction.

>From the point of view of effective field theory \cite{Se97,Fu97} the
models of Refs.\ \cite{Gm92a,Bo89,Su94,Su95} that include up to a
quartic vector self-interaction have the drawback that the
coefficients of some couplings, which should otherwise be present in
the effective Lagrangian truncated at fourth order, have been put
equal to zero without theoretical justification. This fact motivates
us to include the remaining terms and to study their effect on nuclear
matter and finite nuclei. We will show that it is possible to extend
the TM1 set to describe the finite nuclei observables with $Z\geq 8$
and to obtain a description of nuclear matter that follows the DBHF
tendencies better than the conventional non-linear $\sigma-\omega$
models. To do that we will investigate the effects of the new
couplings from EFT keeping the equilibrium properties fixed to those
of TM1. It should be pointed out that, actually, the equilibrium
properties of TM1 lie in the range for which a reasonable description
of finite nuclei properties can be achieved, provided that the EFT
parameters of the model are natural \cite{Fu00}. Here one has to note
that the specific values of TM1 vary from the DBHF result. However,
the RMF sets with only scalar self-interactions which give good
saturation properties deviate sharply from DBHF at high density. Thus,
for our purposes, we believe that a good way to settle the
parametrization is to remain close to the empirical value near
saturation and follow an equation of state similar to that of the DBHF
theory. In the next section we shall detail our strategy.

We emphasize that our goal is not to produce a new optimal set of
parameters intended to compete with well-established conventional sets
like NL3 \cite{Sh93}, which already are very successful for nuclei
both at and away from the line of $\beta$-stability. Instead, we wish
to learn the possibilities of the new EFT models for describing finite
nuclei and for simultaneously tuning the behaviour with density of the
scalar and vector self-energies. In doing this we want to ascertain
whether the comparison with DBHF can provide useful constraints on the
new couplings. For our study, the determination of the parameters of
the model through a least-squares fitting procedure, by calculating
nuclear properties repeatedly until obtaining a best fit, would make
the connections between the resulting parameters and the considered
nuclear observables more obscure.

The paper is organized as follows. Section 2 is devoted to a summary
of the mean field equations and to fit part of the parameters of the
effective Lagrangian to nuclear matter data. We compare our results
with the predictions of other parametrizations available in the
literature. In the third section the remaining parameters of the
effective Lagrangian are obtained by imposing that our mean field
approach reproduces the experimental data for some selected nuclei. A
BCS-type pairing correlation is added in Section 4 to calculate
non-magic even-even nuclei. The extended parameter set is tested in
some applications to nuclear structure phenomena like
isotopic/isotonic energy differences, the isotopic change in the
charge radius and nuclei with large neutron/proton excess. Finally,
the summary and concluding remarks are given in Section 5.

\pagebreak
%
\section{Nuclear matter}
The RMF treatment of the QHD models automatically includes the
spin-orbit force, the finite range and the density dependence of the
nuclear interaction. The RMF model has the advantage that, with the
proper relativistic kinematics and with the meson properties already
known or fixed from the properties of a small number of finite nuclei,
gives excellent results for binding energies, root-mean-square
radii, quadrupole and hexadecapole deformations and other properties
of  spherical and deformed nuclei \cite{Re86,Sh93,Ga90,Ru88,Pa91}. The
quality of the results is comparable to that found in non-relativistic
nuclear structure calculations with effective Skyrme \cite{Va72} or
Gogny \cite{De80} forces.

In recent years the effective field theory approach to QHD has been
studied extensively. The theory and the equations for finite nuclei
and nuclear matter can be found in the literature
\cite{Se97,Fu96,Fu97,Se96} and we shall only outline the formalism
here. We start from Ref.\ \cite{Fu96} where the field equations were
derived from an energy density functional containing Dirac baryons and
classical scalar and vector mesons. Although this energy functional
can be obtained from the effective Lagrangian in the Hartree
approximation \cite{Se97,Fu97}, it can also be considered as an
expansion in terms of ratios of the meson fields and their gradients
to the nucleon mass of a general energy density functional that
contains the contributions of correlations within the spirit of
density functional theory.

According to Refs.\ \cite{Se97,Fu97,Se96} this energy density
functional for finite nuclei can be written as
%
\begin{eqnarray}
{\cal E}({r}) & = &  \sum_\alpha \varphi_\alpha^\dagger({r})
\Bigg\{ -i \mbox{\boldmath$\alpha$} \!\cdot\! \mbox{\boldmath$\nabla$}
+ \beta [M - \Phi({r})] + W({r})
+ \frac{1}{2}\tau_3 R({r})
+ \frac{1+\tau_3}{2} A ({r}) 
\nonumber \\[3mm]
& &  
- \frac{i \beta\mbox{\boldmath$\alpha$}}{2M}\!\cdot\!
  \left (f_v \mbox{\boldmath$\nabla$} W({r})
+ \frac{1}{2}f_\rho\tau_3 \mbox{\boldmath$\nabla$} R({r}) \right)
\Bigg\} \varphi_\alpha (r)
\nonumber \\[3mm]
& & \null
+ \left ( \frac{1}{2}
+ \frac{\kappa_3}{3!}\frac{\Phi({r})}{M}
+ \frac{\kappa_4}{4!}\frac{\Phi^2({r})}{M^2}\right )
 \frac{m_s^2}{g_s^2} \Phi^2({r})  -
\frac{\zeta_0}{4!} \frac{1}{ g_v^2 } W^4 ({r})
\nonumber \\[3mm]
& & \null + \frac{1}{2g_s^2}\left( 1 +
\alpha_1\frac{\Phi({r})}{M}\right) \left(
\mbox{\boldmath $\nabla$}\Phi({r})\right)^2
 - \frac{1}{2g_v^2}\left( 1 +\alpha_2\frac{\Phi({r})}{M}\right)
\left( \mbox{\boldmath $\nabla$} W({r})  \right)^2 
\nonumber \\[3mm]
& &  \null - \frac{1}{2}\left(1 + \eta_1 \frac{\Phi({r})}{M} +
\frac{\eta_2}{2} \frac{\Phi^2 ({r})}{M^2} \right)
 \frac{m_v^2}{g_v^2} W^2 ({r}) 
 - \frac{1}{2e^2} \left( \mbox{\boldmath $\nabla$} A({r})\right)^2
\nonumber \\[3mm]
& & \null
- \frac{1}{2g_\rho^2} \left( \mbox{\boldmath $\nabla$} R({r})\right)^2
- \frac{1}{2} \left( 1 + \eta_\rho \frac{\Phi({r})}{M} \right)
\frac{m_\rho^2}{g_\rho^2} R^2({r}) \,,
\label{eqFN1}
\end{eqnarray}
where the index $\alpha$ runs over all occupied states of the positive
energy spectrum, $\Phi \equiv g_s \phi_0$, $ W \equiv g_v V_0$, $R
\equiv g_\rho b_0$ and $A \equiv e A_0$. Except for the terms with
$\alpha_1$ and $\alpha_2$, the functional (\ref{eqFN1}) is of fourth
order in the expansion. Following Refs.\ \cite{Se97,Fu97,Es99}, we
retain the fifth-order terms $\alpha_1$ and $\alpha_2$ because their
contribution to the nuclear surface energy is numerically of the same
magnitude as the contribution from the quartic scalar term. One can
see that the new terms concentrate on the isoscalar channel and that
the expansion with respect to the isovector meson is shorter (the
$\eta_\rho$ coupling is of third order). Higher non-linear couplings
of the $\rho$ meson are not considered because the expectation value
of the $\rho$ field is typically an order of magnitude smaller than
that of the $\omega$ field \cite{Se97,Fu97}, and they only have a
marginal impact on the usual properties studied for terrestrial
nuclei. For example, in calculations of the high-density equation of
state, M\"uller and Serot \cite{Se96} found the effects of a quartic
$\rho$ meson coupling ($R^4$) to be appreciable only in stars made of
pure neutron matter. A surface contribution $-\alpha_3 \Phi \, (
\mbox{\boldmath$\nabla$} R )^2 /(2 g_\rho^2 M)$ was tested in Ref.\
\cite{Es99} and it was found to have absolutely negligible effects.
We should note, nevertheless, that very recently it has been shown
that couplings of the type $\Phi^2 R^2$ and $W^2 R^2$ are useful to
modify the neutron radius in heavy nuclei while making very small
changes to the proton radius and the binding energy \cite{Ho00}.

The Dirac equation corresponding to the energy density (\ref{eqFN1})
becomes 
\begin{eqnarray}
\Bigg\{-i \mbox{\boldmath$\alpha$} \!\cdot\! \mbox{\boldmath$\nabla$}
& + & \beta [M - \Phi(r)] + W(r)
  +\frac{1}{2} \tau_3 R(r) + \frac{1 +\tau_3}{2}A(r)
 \nonumber \\[3mm]
& & \null  \left.
-  \frac{i\beta \mbox{\boldmath$\alpha$}}{2M}
\!\cdot\! \left [ f_v \mbox{\boldmath$\nabla$} W(r)
 + \frac{1}{2}f_{\rho} \tau_3 \mbox{\boldmath$\nabla$} R(r) \right]
 \right\} \varphi_\alpha (r) =
\varepsilon_\alpha \, \varphi_\alpha (r) \,.
\label{eqFN12}
\end{eqnarray}
The mean field equations for $\Phi$, $W$, $R$ and $A$ are given by
%
\begin{eqnarray}
   -\Delta \Phi(r) + m_s^2 \Phi(r)  & = &
     g_s^2 \rho_s(r)
        -{m_s^2\over M}\Phi^2 (r)
         \left({\kappa_3\over 2}+{\kappa_4\over 3!}{\Phi(r)\over M}
               \right )
                \nonumber  \\[3mm]
     & & \null
       +{g_s^2 \over 2M}
         \left(\eta_1+\eta_2{\Phi(r)\over M}\right)
                { m_v^2\over  g_v^2} W^2 (r)
       +{\eta_{\rho} \over 2M}{g_s^2 \over g{_\rho}^2}
                { m_\rho^2 } R^2 (r)
          \nonumber  \\[3mm]
     & & \null
       +{\alpha_1 \over 2M}[
             (\mbox{\boldmath $\nabla$}\Phi(r))^2
             +2\Phi(r)\Delta \Phi(r) ]
        + {\alpha_2 \over 2M} {g_s^2\over g_v^2}
        (\mbox{\boldmath $\nabla$}W(r))^2 \,,
 \label{eqFN2}  \\[3mm]
 -\Delta W(r) +  m_v^2 W(r)  & = &
 g_v^2 \left( \rho(r) + \frac{f_v}{2} \rho_{\rm T}(r) \right)
 -\left( \eta_1+{\eta_2\over 2}{\Phi(r)\over M} \right ){\Phi(r)
 \over M} m_v^2 W(r)      
   \nonumber  \\[3mm]
         & & \null
      -{1\over 3!}\zeta_0 W^3(r)
          +{\alpha_2 \over M} [\mbox{\boldmath $\nabla$}\Phi(r)
\cdot\mbox{\boldmath $\nabla$}W(r)
                    +\Phi(r)\Delta W(r)] \,,
\label{eqFN3}  \\[3mm]
 -\Delta R(r) +  m_{\rho}^2 R(r)  & = &
     {1 \over 2 }g_{\rho}^2 \left (\rho_{3}(r) + 
     {1 \over 2 }f_{\rho}\rho_{\rm T,3}(r)
  \right ) -  \eta_\rho {\Phi (r) \over M }m_{\rho}^2 R(r) \,,
\label{eqFN4}  \\[3mm]
 -\Delta A(r)   & = &
     e^2 \rho_{\rm p}(r)    \,,
\label{eqFN5}  
\end{eqnarray}
where the baryon, scalar, isovector, proton and tensor densities are
%
\begin{eqnarray}
 \rho(r) & = &
 \sum_\alpha \varphi_\alpha^\dagger(r) \varphi_\alpha(r) \,,
\label{eqFN6} \\[3mm]
 \rho_s(r) & = &
 \sum_\alpha \varphi_\alpha^\dagger(r) \beta \varphi_\alpha(r) \,,
\label{eqFN7} \\[3mm]
 \rho_3 (r) & = &
 \sum_\alpha \varphi_\alpha^\dagger(r) \tau_3 \varphi_\alpha(r) \,,
\label{eqFN8} \\[3mm]
 \rho_{\rm p}(r) & = &
 \sum_\alpha \varphi_\alpha^\dagger(r) \left (\frac{1 +\tau_3}{2} 
 \right)  \varphi_\alpha(r) \,,
\label{eqFN9}  \\[3mm]
 \rho_{\rm T}(r) & = &
 \sum_\alpha \frac{i}{M} \mbox{\boldmath$\nabla$} \!\cdot\!
 \left[ \varphi_\alpha^\dagger(r) \beta \mbox{\boldmath$\alpha$}
      \varphi_\alpha(r) \right] \,,
\label{eqFN10} \\[3mm]
 \rho_{\rm T,3}(r) & = &
 \sum_\alpha \frac{i}{M} \mbox{\boldmath$\nabla$} \!\cdot\!
 \left[ \varphi_\alpha^\dagger(r) \beta \mbox{\boldmath$\alpha$} 
 \tau_3      \varphi_\alpha(r) \right] \,.
\label{eqFN11}  
\end{eqnarray}
In the context of density functional theory it is possible to
parametrize the exchange and correlation effects through local
potentials (Kohn--Sham potentials), as long as those contributions be
small enough that can be considered as minor perturbations to the
potentials \cite{Ko65}. As it is known, this is the case with the
local meson fields. The Hartree values are the ones that control the
dynamics in the relativistic DBHF calculations. Therefore, the meson
fields can also be interpreted as Kohn--Sham potentials. Equations
(\ref{eqFN2})--(\ref{eqFN5}) thus correspond to the Kohn--Sham
equations in the relativistic case \cite{Sp92} and in this sense they
include effects beyond the Hartree approach through the non-linear
couplings \cite{Se97,Fu96,Fu97}.

For infinite nuclear matter all of the gradients of the fields in
Eqs.\ (\ref{eqFN1})--(\ref{eqFN5}) vanish and only the $\kappa_3$,
$\kappa_4$, $\eta_1$, $\eta_2$ and $\zeta_0$ non-linear couplings
remain. Due to the fact that the solution of symmetric nuclear matter
in mean field depends on the ratios $g_s^2/m_s^2$ and $g_v^2/m_v^2$
\cite{Se86}, we have seven unknown parameters. By imposing the values
of the saturation density, total energy, incompressibility modulus and
effective mass, we still have three free parameters (the value of
$g_\rho^2/m_\rho^2$ is fixed from the bulk symmetry energy coefficient
$J$). 

A possible starting point for our study of the effects of the new
terms in the EFT energy density, as mentioned in the Introduction, is
the TM1 parametrization \cite{Su94}. First, because it nicely agrees
with the DBHF calculations with the Bonn-A potential \cite{Br90} for a
wide range of densities. And second, because it provides good results
when applied to finite nuclei calculations, even far away from the
$\beta$-stability line. Our aim is to study the effects of the new
couplings in the description of nuclear matter and finite nuclei and,
at the same time, to improve the TM1 parametrization. Then, instead of
determining the whole set of parameters by a least-squares fit, we
will follow a step-by-step strategy, similar to the one used to
determine the parameter sets in Refs. \cite{serot86,boussy84} in the
relativistic framework, or to determine the Skyrme SkM* parametrization
\cite{bart82} in the non-relativistic case.

According to this strategy we first fix the saturation properties to
be those of TM1 and then introduce the coupling $\zeta_0$ and the new
scalar-vector non-linear couplings $\eta_1$ and $\eta_2$. This way we
can make sure, broadly speaking, that we have the same behaviour of
the equation of state around the saturation point as with TM1. The
addition of the couplings $\zeta_0$, $\eta_1$ and $\eta_2$ cannot be
done with complete freedom once the saturation properties have been
fixed. Including these extra couplings is translated into a
modification of the other coefficients which, eventually, may be
driven to non-natural values. An enlarged discussion of this effect
can be found in Ref.\ \cite{Es99}. To keep all the coefficients within
natural values we find that $\zeta_0=3.6$, $\eta_{1}=1.1$ and
$\eta_{2}=0.1$ is a good choice. It furthermore produces an equation
of state and self-energies in better agreement with DBHF than TM1 (and
also contributes to improve the results for $^{16}$O that is one of
the weak points of TM1, see Section 3). The values of the coupling
constants along with the saturation properties are collected in Table
1. We have denoted this set of parameters as TM1*. We can see that
$\kappa_4$ is positive and that all the coefficients are natural,
i.e., ${\mathcal O}(1)$. The fact that $\kappa_4$ takes a positive
value is very gratifying. Otherwise the energy spectrum has no lower
bound and instabilities in calculations of the equation of state and
of finite systems may occur \cite{Bay60}.

Figure 1 displays the scalar $U_s$ and vector $U_v$ potentials as a
function of the nuclear matter density calculated with TM1*, TM1 (that
contains a quartic vector self-interaction but not the new couplings)
and with the generalized sets G1 and G2 of Ref.\ \cite{Fu97}
(effective field theory model), in comparison with the DBHF result. We
also show the results obtained with the NL3 parameter set \cite{Sh93}
that we have chosen as a representative of the usual non-linear
$\sigma-\omega$ parametrizations with only scalar self-interactions.
Note that the $\kappa_4$ term of NL3 bears a negative sign (Table 1).
Figure 2 shows the equation of state for the different approaches. The
DBHF predictions are believed to be realistic up to a density,
typically, around twice the saturation density \cite{Su94}.

>From Figures 1 and 2 it is clear that the cubic and quartic
self-interactions play a crucial role in following the DBHF results at
high density. The standard non-linear $\sigma-\omega$ sets such as NL3
completely fail in doing so: the vector potential grows almost like a
straight line and gives a much too stiff equation of state. The
quartic vector self-interaction brings down the vector potential and
makes the equation of state softer. This softening of the high-density
equation of state is needed to be consistent with the observed neutron
star masses \cite{Se96}. By construction TM1* gives the same
saturation properties as TM1. However, including the meson
interactions $\eta_1$ and $\eta_2$ we have been able to reproduce the
DBHF results with TM1* better than with TM1, for moderate and high
densities. We have checked that if one tries to reproduce the DBHF
results setting $\eta_1= \eta_2= 0$ this favours large non-natural
values of $\zeta_0$. If we set $\eta_1= \eta_2= 0$ and $\zeta_0$ is
small (roughly $<2$) then $\kappa_4$ remains negative. Only by
introducing the extra constants $\eta_1$ and $\eta_2$ one can agree
better with DBHF, have $\kappa_4>0$ and a not very large $\zeta_0$
value. Nevertheless, we have found that the contributions of the
third-order term $\eta_1 \Phi W^2$ and of the fourth-order term
$\zeta_0 W^4$ in the energy density are far more important than the
contribution of the quartic term $\eta_2 \Phi^2 W^2$. In fact, we
underline that in our calculation $\eta_2$ is compatible with a
vanishing value, indicating that the comparison with DBHF serves to
fix only two couplings of the triad ($\zeta_0, \eta_1, \eta_2$). On
the other hand, the generalized G1 and G2 sets (that were obtained by
fitting finite nuclei observables different from the ones used in TM1)
also show a good agreement with the DBHF results. From Figures 1 and 2
one can see that the results obtained with G1 are similar to those of
TM1, while the predictions of G2 are closer to DBHF\@.

\pagebreak
%
\section{Finite nuclei}
In finite nuclei the contributions from the couplings $\alpha_1$ and
$\alpha_2$ between the scalar field and the gradients of the vector
and scalar fields, as well as the tensor couplings $f_v$ and
$f_\rho$ of the $\omega$ and $\rho$ mesons to the nucleon, do not
vanish. Therefore, we have in principle four more parameters to
adjust, plus the masses of the $\sigma$, $\omega$ and $\rho$ mesons
(or, equivalently, the coupling constants $g_s$, $g_v$ and
$g_\rho$). In accordance with our strategy, we will fix the meson
masses of TM1* to the same values of TM1: $m_s= 511.198$ MeV,
$m_v= 783$ MeV and $m_\rho= 770$ MeV (the nucleon mass is $M= 938$
MeV). In this way we do not mask the influence of the terms that we
want to study.

In our numerical calculation of finite nuclei we have transformed the
Dirac equation (\ref{eqFN12}) into a Schr\"odinger-like equation by
eliminating the small component of the wave function. This equation is
solved by using a standard code for non-relativistic
Skyrme--Hartree--Fock  calculations \cite{Va72}. In the calculations
performed  with the improved TM1 set (TM1*) we use the same
center-of-mass  correction for energies and charge radii as Sugahara
and Toki \cite{Su94} used for TM1:
\begin{equation}
 E_{\rm CM} = \frac{3}{4} \hbar \omega , \qquad
 r^2_{\rm ch}=r^2_{\rm p}+0.64-{3\over 2} \left({b^2\over A}\right)
 \hspace{2mm} \rm{fm^2} ,
\end{equation}
where $\hbar\omega = 41 A^{-1/3}$ MeV, $b=\sqrt{\hbar/m\omega}$ is the
harmonic oscillator parameter and 0.64 fm$^2$ takes into account the
finite size correction of the proton \cite{negele70}.

We obtain the coupling constants $\alpha_1$, $\alpha_2$, $f_v$,
$\zeta_0$, $\eta_1$ and $\eta_2$ (the last three combined with the
nuclear matter calculation as explained in Section 2) by imposing that
the total energy, the charge radius and the 1$p$ splitting for
neutrons and protons of the symmetric nucleus $^{16}$O be as close as
possible to the experimental values. To deal with asymmetric nuclei
the $g_{\rho}$, $\eta_{\rho}$ and $f_{\rho}$ couplings are needed.
Following Ref.\ \cite{Su94} we fix the volume asymmetry coefficient
\begin{equation}
J={k^2_F\over{6(k^2_F+{M^*_\infty}^2)^{1/2}}}
+{g^2_{\rho}k_F^2\over
{12\pi^2m_{\rho}^2}}{1\over{1+\eta_{\rho}(1-M^{*}_{\infty}/M)}}
\end{equation}
to be 36.9 MeV\@. Actually this value corresponds to the difference
between the neutron and nuclear matter DBHF energies per particle
\cite{li92} calculated at the nuclear matter saturation density, which
is known to be a good approach for estimating $J$ \cite{poll99}. Then
we determine $\eta_\rho$ and $f_{\rho}$ so that the energy of
$^{208}$Pb be the experimental one. The tensor coupling $f_\rho$
happens  to be useless in our fitting: its contribution, as previously
reported \cite{Es99,Ru88}, is negligible and we have taken $f_\rho= 0$
for TM1*. This is not the case for the coupling $\eta_\rho$, whose
influence is noticeable \cite{Es99}. As a final step in our fitting
procedure we have to check that the values of all the parameters are
natural. The whole set of parameters of TM1* is given in Table 1. 

We should like to discuss the systematics of the finite nuclei
properties with the new couplings in some more detail. The bulk
parameters ($\zeta_0, \eta_1, \eta_2$) only have a slight influence on
the binding energies ($B$) and charge radii ($r_{\rm ch}$), and
practically no effect on the spin-orbit splittings ($\Delta E_{\rm
SO}$). The incidence of $\eta_2$ is again negligible compared to
$\zeta_0$ and $\eta_1$. However, if ($\zeta_0, \eta_1, \eta_2$) are
given the wrong values then it may be impossible to correct the
results for $B$ and $r_{\rm ch}$ with only the surface parameters
($f_v, \alpha_1, \alpha_2$). Thus, one first needs a reasonable ansatz
for ($\zeta_0, \eta_1, \eta_2$) to be able to get acceptable values
for $B$ and $r_{\rm ch}$. It was not obvious a priori that the
($\zeta_0, \eta_1, \eta_2$) values favoured by the comparison with
DBHF would fall into this category. If all other parameters are kept
fixed, decreasing $\alpha_1$ makes $B$ and $\Delta E_{\rm SO}$ larger
and $r_{\rm ch}$ smaller (we define $B$ to be positive). The coupling
$\alpha_2$ has just the opposite effect. For the same change in
$\alpha_1$ as in $\alpha_2$, the modifications on $B$, $r_{\rm ch}$
and $\Delta E_{\rm SO}$ are roughly twice larger with $\alpha_1$ than
with $\alpha_2$. Once a set of ($\zeta_0, \eta_1, \eta_2$) values is
specified, $f_v$ serves to bring the strength of $\Delta E_{\rm SO}$
closer to the desired value. Then the couplings $\alpha_1$ and
$\alpha_2$ are used for the fine tuning of the $B$, $r_{\rm ch}$ and
$\Delta E_{\rm SO}$ values. We point out that after specifying
($\zeta_0, \eta_1, \eta_2$) almost the same $B$, $r_{\rm ch}$ and
$\Delta E_{\rm SO}$ are obtained with many distinct families of
($\alpha_1, \alpha_2$) values. In principle, we have realized that
making $\alpha_1$ small or negative and readjusting $\alpha_2$ to
recover the same binding energies and spin-orbit splittings, helps one
to eventually get slightly larger radii. That is, from the interplay
of $\alpha_1$ and $\alpha_2$ it is possible to achieve some change in
the value of $r_{\rm ch}$ relative to the value of $B$, but the effect
is not very significant. We are led to conclude that at least one of
the three couplings ($f_v, \alpha_1, \alpha_2$) is not singled out by
the properties analyzed, and that a correlation exists between these
surface parameters and the bulk parameters ($\zeta_0, \eta_1,
\eta_2$). 

As a first test of the full TM1* parametrization we have calculated
the surface energy coefficient $E_s$ and the surface thickness $t$ of
the density profile (standard 90\%-10\% fall-off distance of the
nuclear density) in semi-infinite nuclear matter. The results are
shown in Table~2. The surface energy obtained with TM1* lies within
the region of empirical values, whereas the surface thickness $t$ is
slightly small \cite{Es99,Ce93c}. The energies, charge radii and
spin-orbit splittings of the magic nuclei $^{16}$O and $^{208}$Pb used
in our fit as well as the values for $^{40}$Ca, $^{48}$Ca and
$^{90}$Zr, which were included in the fit of TM1 \cite{Su94}, are also
displayed in Table 2. We show the experimental values and the results
obtained with the sets TM1*, TM1 and NL3 \cite{Sh93} (non-linear model
with only scalar self-interactions), and with the generalized
parameter sets G1 and G2 of Ref.\ \cite{Fu97}. In addition to the
couplings listed in Table 1 for G1 and G2, these sets have a few more
parameters related with the electromagnetic structure of the pion and
the nucleon (see Ref.\ \cite{Fu97}), which we have taken into account
for Table 2. The TM1 results for $^{16}$O are given here for
completeness, as we recall that TM1 was devised for heavier nuclei
\cite{Su94}. Concerning the influence of the centre-of-mass motion on
the energy and the charge radius, it should be noted that different
parametrizations use, in general, different prescriptions. Due to the
fact that the centre-of-mass corrections are included in the fit of
the parameters, we report in Table 2 the values we have obtained with
the same prescription as the authors used in their original works.

The TM1* calculations for magic nuclei displayed in Table 2 reproduce
the experimental energies within $\sim 0.8\%$ and the charge radii and
spin-orbit splittings with a similar quality to the successful NL3
\cite{Sh93}, G1 or G2 \cite{Fu97} parametrizations. In order to check
the ability of TM1* for describing nuclei far from the stability line,
we have calculated the energy and charge radius of some drip-line
(double-closed shell) nuclei, namely $^{56}$Ni, $^{78}$Ni, $^{100}$Sn
and $^{132}$Sn. Table 2 shows that all the forces considered here
produce similar results for the energy per particle and the charge
radius of finite nuclei, which agree well with experiment. The
single-particle energies of neutrons and protons are compared with the
experimental data in Figures 3a and 3b for the $^{208}$Pb nucleus with
the TM1*, TM1 and NL3 sets. One can see that all these
parametrizations qualitatively describe the experimental values.
Although the nuclear matter properties are equal in TM1 and TM1*, the
spectra are slightly different mainly due to the tensor coupling $f_v$
present in TM1*, which has a noticeable influence in the spin-orbit
potential \cite{Se97,Fu96,Es99,Fu98}.

\pagebreak
%
\section{Even-even nuclei}
To describe even-even nuclei other than double magic nuclei we
introduce the pairing correlation in the BCS approximation with a
constant gap $\Delta$, as in earlier calculations
\cite{Su94,Re86,Ga90}. It is to be kept in mind that the seniority
pairing recipe is not appropriate for exotic nuclei near the drip
lines because the coupling to the continuum is not treated properly.
The fact that continuum states become significantly populated as one
approaches the drip lines can be taken into account in the
relativistic Hartree-plus-Bogoliubov (RHB) method
\cite{egido96,vretenar98,farkan00}. However, it has been pointed out
in Ref.\ \cite{Ch98} that a qualitative estimation of the drip lines
can be obtained within the BCS scheme by taking into account some
quasi-bound states owing to their centrifugal barrier which mocks up
the influence of the continnum.

In order to be as consistent as possible with TM1 here we take the gap
energy $\Delta= 11.2/\sqrt{A}$ MeV, that corresponds to the widely
used phenomenological formula of Bohr and Mottelson \cite{bohr1}. In
practice we have found that the same gap energy is obtained by fitting
the Sn isotopic energy difference \cite{Ch98}
\begin{equation}
\Delta E = [ E - E(^{116}{\rm Sn}) ]_{\rm BCS} - 
[ E - E(^{116} {\rm Sn}) ]_{\rm Exp.}
\label{eqFN22}
\end{equation}
calculated with TM1*. We restrict the number of active shells to the
occupied shells contained in a major harmonic oscillator shell above
and below of the last closed shell. When the nuclei approach the drip
lines there are not bound single-particle levels above the chemical
potential. In this case we take the bound-state contributions as well
as those coming from quasi-bound states at positive energies
\cite{Ch98}. 

In Table 3 we report the energy and charge radii of $^{58}$Ni,
$^{116,124}$Sn and $^{184,196,214}$Pb that were used in the TM1 fit.
TM1* shows an agreement with experiment similar to that found for TM1.
Apart from the results presented in Tables 2 and 3, we have compared
the energy given by TM1* for several light nuclei of $Z\leq 20$ with
the results given by TM2 \cite{Su94} (as TM2 was designed for $Z\leq
20$) and with the TM1 results. TM1* improves the TM1 results in this
region and the quality of the energies is similar to that of TM2. In
the following we will calculate isotopic and isotonic energy
differences, isotopic shifts in charge radii and two-neutron and
two-proton separation energies near and away from the
$\beta$-stability line, to examine whether TM1* is also acceptable for
these properties in comparison with experiment and with other
relativistic sets.

\subsection{Isotopic and isotonic energy differences}
We have calculated the isotopic energy differences $\Delta E$ for
several Sn and Pb isotopes (referred to $^{116}$Sn and $^{208}$Pb,
respectively) with the TM1*, TM1 and NL3 parametrizations. The results
are shown in Figures 4a and 4b. In the case of the Sn isotopes there
are some differences between the NL3 results and those of TM1 or TM1*.
The NL3 isotopic energy differences appreciably deviate from the
experiment for Sn isotopes with a neutron number $N$ larger than 66,
while the TM1 or TM1* results remain close to the experimental values.
If we compare with the non-relativistic calculations performed in
Ref.\ \cite{Ch98} with the Skyrme forces SLy4 and SkM*, the NL3
results qualitatively behave as those of SkM*, whereas the TM1 and
TM1* predictions are closer to those of SLy4. The results for the Pb
isotopes are shown in Figure 4b, and for $N= 82$ isotones (referred to
$^{132}$Sn) in Figure 4c. The TM1*, TM1 and NL3 sets show different
trends for the lead isotopes. TM1 predicts better $\triangle E$ values
over the other parameter sets. For the $N < 126$ isotopes TM1 and TM1*
predict an arch structure similar to the one found with the SLy4
interaction \cite{Ch98}, while NL3 shows a structure more similar to
the SkM* force \cite{Ch98}. For $N > 126$, $\triangle E$ increases as
a function of $N$ for the three relativistic sets, similarly to the
SLy4 calculation \cite{Ch98}.

The $N=82$ isotone energy differences found with NL3, TM1 and TM1*
show a completely different behaviour compared with SLy4 and SkM*. In
the relativistic case $\triangle E$ decreases with increasing $Z$ up
to $Z=56$ and increases afterwards. The largest separation with
respect to the experimental value corresponds to $Z=56$, although the
difference is more pronounced for TM1 and TM1* than for NL3. In the
non-relativistic calculations \cite{Ch98} one finds an arch structure
with SLy4, with the largest difference with experiment corresponding
to $Z=56$ (although it is positive in this case), and a monotonous
increasing of $\triangle E$ as a function of $Z$ with SkM*. This
qualitatively different behaviour in the $N=82$ isotopic chain could
be caused by the different pairing interaction used in the present
calculation (constant gap) and in the non-relativistic calculations of
Ref.\ \cite{Ch98} where a density-dependent zero-range pairing force
(more similar to a constant strength) was considered.

\subsection{Isotopic change in charge radius}
In the past years the isotopic shifts in charge radii have been
studied for the isotopic chain of Pb nuclei using various techniques
\cite{Ch98,tajima93,sharma93}. Non-linear $\sigma-\omega$ calculations
with a constant gap pairing interaction in general reproduce the
experimentally observed kink in the isotopic shifts about $^{208}$Pb
\cite{Sh93,sharma93}. However, the standard non-relativistic (zero
range or finite range) forces are not able to describe this kink. Only
by improving the pairing interaction and by taking into account some
terms usually not considered in the Skyrme functional and the two-body
center-of-mass correction, the non-relativistic results agree with the
experimental observation \cite{Ch98,tajima93}. Here we have calculated
the Pb isotopic shifts with the TM1* parameter set. In Figure~5 the
result is compared with the prediction of the TM1 and NL3 sets, and
also with the experimental data. All these parameter sets yield
qualitatively similar results and reproduce the experimental kink
reasonably well. Notice that these Pb isotopic shifts are not included
in the TM1 and TM1* fits.

\subsection{Two-neutron and two-proton separation energies}
We have evaluated the two-neutron $S_{\rm 2n}$ and two-proton 
$S_{\rm 2p}$ separation energies from the calculated energies using
\cite{bohr1} 
\begin{eqnarray}
S_{\rm 2n}(N,Z) & = & E(N-2,Z)-E(N,Z) ,
\\
S_{\rm 2p}(N,Z) & = & E(N,Z-2)-E(N,Z) .
\end{eqnarray}
The $S_{\rm 2n}$ values for the illustrative cases of $Z= 20$ and 50
as well as the $S_{\rm 2p}$ value for $N=82$ with the TM1*, NL3 and
TM1 sets are presented in Figures 6a, 6b and 6c, respectively. The
experimental data are also given for comparison.

On the wole, the $S_{\rm 2n}$ and $S_{\rm 2p}$ values obtained from
TM1* agree well with the experimental observation and also with the
predictions of TM1 and NL3 (except for a slight discrepancy for some
specific cases). In concrete, for Ca isotopes the shell effects at
$N=20$ and 28 are well reproduced by the three relativistic sets.
Another shell effect is predicted at $N=38$ although no experimental
information is available to confirm it. Something similar happens for
Sn isotopes at $N=50$ and at $N=82$ where the calculated results
qualitatively agree with the experimental value. With respect to the
isotone chain of $N=82$ no experimental information exists to confirm
the shell effect at $Z=50$. In this case the relativistic sets are not
able to quantitatively reproduce the experimental $S_{\rm 2p}$ energies
in the $Z=54$--58 region, due maybe to the adopted pairing scheme. The
$S_{\rm 2n}$ value decreases with increasing the neutron number and
vanishes at the neutron-drip line. Similarly, the $S_{\rm 2p}$ value
decreases with increasing proton number as the proton-drip line is
reached. 

As we have mentioned, although the BCS approach to pairing
correlations is not well suited for dealing with the drip lines
\cite{egido96,vretenar98,farkan00}, an estimate can be given with the
present BCS calculation that takes into account some continuum effects
through the quasi-bound levels \cite{Ch98}. The authors of Ref.\
\cite{Su94} discussed the unability of TM1 for describing Zr isotopes
with $N$ larger than 82, while they could be described with the NL1
parametrization. We have found that to be able to describe these
nuclei, which have a chemical potential close to zero, it is crucial
for the BCS calculation to take into account the quasi-bound levels
$2f_{5/2}$, $1h_{9/2}$ and $1i_{13/2}$ that lie a few MeV above the
Fermi level. In this way we have estimated the neutron-drip line for
Zr at $N\sim 98$. Similarly, we have estimated the neutron-drip line
for Ca, Sn and Pb isotopes at $A\sim$ 62, 164 and 264, respectively,
for all the analysed parameter sets. These BCS estimates are in good
agreement with the results of the non-relativistic interactions SLy4
and SkM* reported in Ref.\ \cite{Ch98}, where almost the same
technique was used for dealing with the pairing correlations. For
$N=82$ isotones we find the proton-drip line at $A\sim 156$, which
corresponds to $^{156}$W in agreement with experimental information
\cite{Page92}. 

\pagebreak

\section{Summary and conclusions}
We have explored whether the parameter set TM1 \cite{Su94} can be
improved by adding new couplings that stem from the modern effective
field theory approach to relativistic nuclear phenomenology. We have
been concerned with analyzing the possibilities of the new couplings
to ensure a reasonable agreement with the density dependence of the
scalar and vector components of the DBHF self-energies, while
performing well for finite nuclei. The extended parameter set has been
called TM1*. It is able to reproduce ground-state properties of
spherical nuclei for $Z\geq 8$ with a quality similar to conventional
sets like NL1 or NL3, and with the appealing feature of having a
positive quartic scalar self-coupling. This could not be achieved with
the set TM1 which had to be restricted to $Z$ larger than 20 in order
to keep $\kappa_4$ positive \cite{Su94}. It is important to note that
this limitation seems to be common to any set of parameters containing
only a quartic vector self-interaction on top of the standard
non-linear $\sigma-\omega$ model. To check this point we have
performed calculations with the recently proposed NL-SV1 and NL-SV2
parameter sets \cite{farkan00} that include a quartic vector
self-coupling (like TM1). For light nuclei we find a good agreement
with experiment when we use the NL-SV1 set which has a negative
$\kappa_4$ coupling, whereas this is not the case with the NL-SV2 set
where $\kappa_4$ is positive.

In comparison with the DBHF results in nuclear matter the extended set
TM1* shows a significant improvement over TM1 due to the addition of
the $\eta_1$ and $\eta_2$ couplings. The latter couplings (at least
$\eta_1$) are very helpful to bring the vector and scalar potentials
closer towards the DBHF calculations as the density grows. To the end
of computing finite nuclei we have introduced the $f_v$,
$\alpha_1$, $\alpha_2$, $\eta_\rho$ and $f_\rho$ parameters on top of
the set that describes nuclear matter. We remark that the new
parameters have a minor influence on the investigated properties of
finite nuclei. However, they allow the full TM1* force to improve the
agreement with experiment for double-closed shell nuclei compared with
the starting TM1 parameters and to obtain better results for
light-mass nuclei, which was a shortcoming of the TM1 set. We also
have tested the TM1* force for isotopic energy differences, isotopic
changes in charge radii and two-neutron and two-proton separation
energies. Nuclei near the drip lines have been explored for some
particular cases by taking into account quasi-bound states in the BCS
calculation following the method of Ref.\ \cite{Ch98}. It should be
mentioned that by including all of the relevant couplings in the
energy density expansion compatible with the EFT approach to QHD, as
developed in Refs.\ \cite{Se97,Fu97}, the TM1* model is more
consistent with our current understanding of effective field theories.
Nevertheless, we have seen that some of the new couplings of the EFT
model remain underdetermined in spite of the information taken into
account about the equation of state and the self-energies at higher
densities. 

In conclusion, the relativistic mean field approach extended by the
new non-linear meson self-interactions and tensor couplings based upon
effective field theory, allows one to reproduce at the same time the
trends of microscopic DBHF calculations up to relatively high
densities and various finite nuclei properties. In the low-density
domain (that corresponds to the finite nuclei region) the main
properties are almost fixed by the nuclear matter properties around
saturation, and then the new parameters have only a small
contribution. However, as the density increases the vector-vector and
scalar-vector meson interactions play an important role in providing
enough flexibility to the model to be able to follow the tendency of
the DBHF calculations. Extended sets like TM1* may be more useful for
systems having relatively higher density and temperature, whereas they
will serve the same purpose for normal systems as the conventional
parameter sets. To further constrain the new EFT parameters additional
observables will be required. Nuclear phenomena involving currents
could prove helpful for couplings such as $\alpha_1$ and $\alpha_2$
that imply the derivatives of the fields. On the side of the isovector
channel, information from many-body DBHF calculations of asymmetric
and neutron matter as well as data on neutron radii and the neutron
skin thickness should be relevant.

\section{Acknowledgements}
The authors would like to acknowledge support from the DGICYT (Spain)
under grant PB98-1247 and from DGR (Catalonia) under grant
1998SGR-00011.  M.D.E. acknowledges financial support from the CIRIT
(Catalonia). S.K.P. thanks the Spanish Education Ministry grant
SB97-OL174874 for financial support and the Departament d'Estructura i
Constituents de la Mat\`eria of the University of Barcelona for kind
hospitality. 

\pagebreak

%

%

\pagebreak

%
%
%
\section*{Figure captions}
\begin{description}
\item[Figure 1.]
Scalar $U_s$ and vector $U_v$ potentials against the nuclear matter
density as obtained in a DBHF calculation with the Bonn-A potential
\cite{Br90}  and with the relativistic mean field parametrizations
TM1*, TM1 \cite{Su94}, G1 and G2 \cite{Fu97}.
\item[Figure 2.]
    Equation of state for the same cases as in Figure 1.
\item[Figure 3.]
The single-particle energies for $^{208}$Pb obtained by various
relativistic mean-field parametrizations are compared with the
experimental  data for neutrons (a) and protons (b).

\item[Figure 4.]
The isotopic energy difference obtained with the TM1* parameter set is
compared with the TM1 and NL3 calculations for Sn isotopes (a) and Pb
isotopes (b). Plot (c) shows the isotonic energy difference for $N=82$.

\item[Figure 5.]
The isotopic shifts in charge radii for the $Z=82$ chain.

\item[Figure 6.]
The calculated separation energies are compared with the experimental
data: (a) two-neutron separation energy $S_{\rm 2n}$ for $Z=20$, (b)
two-neutron separation energy for $Z=50$, and (c) two-proton
separation energy $S_{\rm 2p}$ for $N=82$.

\end{description}

\pagebreak

%
\section*{Tables}
\vspace*{1cm}

\begin{table}[h]
\caption{Various parameter sets for the relativistic energy density 
functional and the corresponding saturation properties. The 
coupling constants are dimensionless. }
\vspace*{1.cm}
\centering
\begin{tabular}{ccccccccccc}
\hline
\hline
  & & TM1*  & &TM1 & & NL3 && G1 && G2 \\
\hline
$m_s/M$      & &  0.545  & &  0.545  & & 0.541 & & 0.540 & & 0.554\\
$g_s/4\pi$   & & 0.893 & &  0.798 & &0.813 & & 0.785 & & 0.835\\
$g_v/4\pi$   & &  1.192 & &   1.003 & & 1.024 & & 0.9650 & & 1.016 \\
$g_\rho/4\pi$   & &  0.796 & &   0.737 & & 0.712 & & 0.698 & & 0.755\\
$\kappa_3$  & & 2.513 & &  1.021 & & 1.465 & & 2.207 & & 3.247 \\
$\kappa_4$ & & 8.970 & & 0.124 & & $-$5.668 & & $-$10.090 & & 0.632\\
$\zeta_0$ & &  3.600 & &  2.689 & &  0.0 & & 3.525 & & 2.642\\
$\eta_1$ & &  1.1 & & 0.0 & & 0.0 & & 0.071 & & 0.650\\
$\eta_2$ & &  0.1 & & 0.0  & & 0.0 & & $-$0.962 & & 0.110\\
$\alpha_1$ & &  $-$0.15 & & 0.0  & & 0.0 & & 1.855 & & 1.723\\
$\alpha_2$ & & $-$2.20 & & 0.0 & & 0.0 & & 1.788 & & $-$1.580\\
$f_v/4$ & &  0.06 & & 0.0 & & 0.0 & & 0.108 & & 0.173 \\
$\eta_\rho$ & & 0.45 & & 0.0 & & 0.0 & & $-$0.272 & & 0.390\\
\hline
$a_v$ (MeV)  & & $-$16.30 &  &$-$16.30 & & $-$16.24 & & $-$16.14 & & $-$16.07\\
$\rho_\infty$ (fm$^{-3}$) & & 0.145&  & 0.145& & 0.148 & & 0.153 & & 0.153\\
$K$ (MeV)     &   & 281.1 & & 281.1 & &271.5 & & 215.0 & & 215.0\\
$M^{*}_{\infty}/M$  & & 0.634  & & 0.634 & & 0.595 & & 0.634 & & 0.664\\
$J$ (MeV)  & &  36.90   & & 36.90 & & 37.40 & & 38.5 & & 36.4\\
\hline
\hline
\end{tabular}
\end{table}

\pagebreak

\begin{table}
\vspace*{-1.3cm}
\caption{The surface energy coefficient $E_s$, surface thickness
$t$,  energy per nucleon $E/A$, charge radius $r_{\rm ch}$ and
spin-orbit  splittings $\Delta E_{\rm SO}$ of the least-bound nucleons
using the TM1*, TM1, NL3, G1 and G2 parameter sets are compared with
the experimental data. The energies are given in MeV, while $t$ and
$r_{\rm ch}$ are given in fm. The experimental values of $E/A$ for
$^{78}$Ni and $^{100}$Sn are, in fact, extrapolated data \cite{Ch98}.}
\vspace*{0.6cm}
\centering
\small
\begin{tabular}{lclrrrrrrrrrrrr}
\hline
\hline
    & & & & TM1* & & TM1 & & NL3 & & G1 & & G2  & & Exp. \\
\hline
&& $E_s$    & &
  18.57  & & 18.51 & & 18.36  & & 18.06 & & 17.80 & & 16.5--21.0\\
&& $t$   & &
  1.90  & & 1.91 & & 1.99     & & 1.98 &  & 2.08 & & 2.2--2.5\\
\hline
$^{16}$O & &
  $E/A$  & & $-$8.02 & & $-$8.15 & & $-$8.08 & & $-$7.97 & & $-$7.97
& &  $-$7.98\\
& & $r_{\rm ch}$ & & 2.67  & & 2.66 & & 2.73 & & 2.72 
& & 2.72 & & 2.73 \\
& & $\Delta E_{\rm SO}$ (n,$1p$) & & 6.3 & & 5.6 & & 6.4 & &6.0 
& & 5.9 & & 6.2 \\
& & \hspace{1.1cm} (p,$1p$) & & 6.2 & &  5.6 & & 6.3 & & 5.9 & & 5.9
& & 6.3 \\
$^{40}$Ca & &
  $E/A$  & & $-$8.55 & & $-$8.62 & & $-$8.54 & & $-$8.55 & & $-$8.55
& &  $-$8.55\\
& & $r_{\rm ch}$ & & 3.44  & & 3.44 & & 3.48   & & 3.46   
& & 3.45  & & 3.48\\
& & $\Delta E_{\rm SO}$ (n,$1d$) & & 6.3 & & 5.7 & & 6.7 & & 6.6 
& & 6.5 & & 6.3\\
& & \hspace{1.1cm} (p,$1d$) & & 6.3 & &  5.7 & & 6.6  & &  6.5 & & 6.4
& & 7.2\\
$^{48}$Ca & &
  $E/A$  & & $-$8.64 & & $-$8.65 & & $-$8.64 & & $-$8.67 & & $-$8.68
& &  $-$8.67 \\
& & $r_{\rm ch}$ & & 3.46  & & 3.46 & & 3.48 & & 3.44 & & 3.44  
& & 3.47 \\
& & $\Delta E_{\rm SO}$ (n,$1d$) & & 5.4 & & 5.0 & & 6.1 & & 5.8 
& & 5.6 & & 3.6\\
& & \hspace{1.1cm} (p,$1d$) & & 5.6 & &  5.2 & & 6.3  & &  6.2 & & 6.0
 & & 4.3\\
$^{90}$Zr & &
  $E/A$  & & $-$8.72 & & $-$8.71 & & $-$8.69 & & $-$8.71 & & $-$8.68
& &  $-$8.71\\
& & $r_{\rm ch}$ & & 4.26  & & 4.27 & & 4.28 & & 4.28 & & 4.28 
& & 4.26 \\
& & $\Delta E_{\rm SO}$ (n,$2p$) & & 1.6 & & 1.4 & & 1.6 & &1.8 
& & 1.8 & & 0.5 \\
$^{208}$Pb & &
  $E/A$  & & $-$7.87 & & $-$7.87 & & $-$7.87 & & $-$7.87 & & $-$7.86
& &  $-$7.87\\
& & $r_{\rm ch}$ & & 5.53  & & 5.54 & & 5.52   & & 5.50 & & 5.50 
& & 5.50\\
& & $\Delta E_{\rm SO}$ (n,$3p$) & & 0.8 & & 0.7 & & 0.8 & & 0.9 
& & 0.9
& & 0.9\\
& & \hspace{1.1cm} (p,$2d$) & & 1.7 & &  1.4 & & 1.6  & &  1.8 & & 1.8
 & & 1.3\\
$^{56}$Ni & &
  $E/A$  & & $-$8.60 & & $-$8.56 & & $-$8.60 & & $-$8.61 & & $-$8.60
& & $-$8.64\\
& & $r_{\rm ch}$ & & 3.74  & & 3.74 & & 3.72   & & 3.72 & & 3.73 
& & 3.76\\
$^{78}$Ni & &
  $E/A$  & & $-$8.18 & & $-$8.19 & & $-$8.23 & & $-$8.28 & & $-$8.28
& &  $-$8.23\\
& & $r_{\rm ch}$ & & 3.95  & & 3.95 & & 3.95 & & 3.92 & & 3.92 
& & -- \\
$^{100}$Sn & &
  $E/A$  & & $-$8.30 & & $-$8.27 & & $-$8.28 & & $-$8.28 & & $-$8.27
& &  $-$8.26\\
& & $r_{\rm ch}$ & & 4.49  & & 4.49 & & 4.48   & & 4.47   & & 4.47 
& & --\\
$^{132}$Sn & &
  $E/A$  & & $-$8.33 & & $-$8.34 & & $-$8.36 & & $-$8.38 & & $-$8.37
& &  $-$8.35 \\
& & $r_{\rm ch}$ & & 4.72  & & 4.73 & & 4.72   & & 4.69   & & 4.69 
 & & --\\
\hline
\hline
\end{tabular}
\end{table}

\pagebreak

\begin{table}
\caption{Same as Table 2 for some open shell nuclei.}
\vspace*{1.cm}
\centering
\begin{tabular}{lclrrrrrrrrrrrr}
\hline
\hline
    & & & & TM1* & & TM1 & & NL3 & & G1 & & G2  & & Exp. \\
\hline
\hline

$^{58}$Ni & &
  $E/A$  & & $-$8.64 & & $-$8.61 & & $-$8.63 & & $-$8.62& & $-$8.62
& & $-$8.73\\
& & $r_{\rm ch}$ & & 3.76  & & 3.76 & & 3.75   & & 3.74 & & 3.75
& & 3.77\\
$^{116}$Sn & &
  $E/A$  & & $-$8.52 & & $-$8.52 & & $-$8.49 & & $-$8.48& & $-$8.48
& &  $-$8.52\\
& & $r_{\rm ch}$ & & 4.62  & & 4.62 & & 4.62   & & 4.60   & & 4.60 
& & 4.63\\
$^{124}$Sn & &
  $E/A$  & & $-$8.45 & & $-$8.46 & & $-$8.45 & & $-$8.46 & & $-$8.45
& &  $-$8.47 \\
& & $r_{\rm ch}$ & & 4.67  & & 4.67 & & 4.67   & & 4.65   & & 4.64 
 & &4.67\\
$^{184}$Pb & &
  $E/A$  & & $-$7.81 & & $-$7.80 & & $-$7.77 & & $-$7.75 & & $-$7.74
& &  $-$7.78\\
& & $r_{\rm ch}$ & & 5.41  & & 5.41 & & 5.40   & & 5.39 & & 5.38 
& & --\\
$^{196}$Pb & &
  $E/A$  & & $-$7.89 & & $-$7.87 & & $-$7.86 & & $-$7.85 & & $-$7.84
& &  $-$7.87\\
& & $r_{\rm ch}$ & & 5.47  & & 5.48 & & 5.46   & & 5.45 & & 5.44 
& & --\\
$^{214}$Pb & &
  $E/A$  & & $-$7.75 & & $-$7.76 & & $-$7.75 & & $-$7.74 & & $-$7.73
& &  $-$7.77\\
& & $r_{\rm ch}$ & & 5.59  & & 5.59 & & 5.58   & & 5.55 & & 5.54 
& & --\\
\hline
\hline
\end{tabular}
\end{table}


\begin{thebibliography}{99}
%
\parskip= -1.5mm
%
\bibitem{Ma89} R. Machleidt, Adv.\ Nucl.\ Phys.\ {\bf 19}, 189 (1989).
\bibitem{Ho84} C. J. Horowitz and B. D. Serot, 
               Phys.\ Lett.\ B {\bf 137}, 283 (1984).
\bibitem{An83} M. R. Anastasio, L. S. Celenza, W. S. Pong, 
               and C. M. Shakin, Phys.\ Rep.\ {\bf 100}, 327 (1983).
\bibitem{Br84} R. Brockmann and R. Machleidt, 
               Phys.\ Lett.\ B {\bf 149}, 283 (1984).
\bibitem{Ha87} B. ter Haar and R. Malfliet, 
               Phys.\ Rep.\ {\bf 149}, 207 (1987).
%
\bibitem{Se86} B. D. Serot and J. D. Walecka,  
Adv.\ Nucl.\ Phys.\ {\bf 16}, 1 (1986).
%
\bibitem{Se97} B. D. Serot and J. D. Walecka, 
Int.\ J. of Mod.\ Phys.\  E {\bf 6}, 515 (1997).
%
\bibitem{Wa74} J. D. Walecka,  
Ann.\ Phys.\ (N.Y.) {\bf 83}, 491 (1974).
%
\bibitem{Bo77} J. Boguta and A. R. Bodmer, 
Nucl.\ Phys.\ {\bf A292}, 413 (1977).
%
\bibitem{Fu96} R. J. Furnstahl, B. D. Serot, and H. B. Tang,
               Nucl.\ Phys.\ {\bf A598}, 539 (1996).
%
\bibitem{Fu97} R. J. Furnstahl, B. D. Serot, and H. B. Tang,
                Nucl.\ Phys.\ {\bf A615}, 441 (1997).
%
\bibitem{Se96} H. M\"uller and B. D. Serot, 
Nucl.\ Phys.\ {\bf A606}, 508 (1996).
%
\bibitem{Fu97b} J. J. Rusnak and R. J. Furnstahl, 
                Nucl.\ Phys.\ {\bf A627}, 495 (1997).
%
\bibitem{Cl98} B. C. Clark, R. J. Furnstahl, L. K. Kerr, J. Rusnak,
               and S. Hama, Phys.\ Lett.\ B {\bf 427}, 231 (1998).
%
\bibitem{Fu00} R. J. Furnstahl and B. D. Serot, 
               Nucl.\ Phys.\ {\bf A671}, 447 (2000).
%
\bibitem{Es99} M. Del Estal, M. Centelles, and X. Vi\~nas,
               Nucl.\ Phys.\ {\bf A650}, 443 (1999).
%
\bibitem{Fu98} R. J. Furnstahl,  J. J. Rusnak, and B. D. Serot,
               Nucl.\ Phys.\ {\bf A632}, 607 (1998).
%
\bibitem{Wa00} P. Wang, Phys.\ Rev.\ C {\bf 61}, 054904 (2000).
%
%
\bibitem{Co70} F. Coester, S. Cohen, B. D. Day, and C. M. Vincent,
                    Phys.\ Rev.\ C {\bf 1} 769 (1970).
%
\bibitem{Am92} A. Amorim and J. A. Tjon, 
Phys.\ Rev.\ Lett.\ {\bf 68}, 772 (1992).
%
\bibitem{Gm91} S. Gmuca, J. Phys.\  G {\bf 17}, 1115 (1991).
%
\bibitem{rashdan97} M. Rashdan, Phys.\ Lett.\ B {\bf 395}, 141 (1997).
%
\bibitem{Gm92a} S. Gmuca, Z. Phys.\  A {\bf 342}, 387 (1992); 
                          Nucl.\ Phys.\ {\bf A547}, 447 (1992). 
%
\bibitem{Bo89} A. R. Bodmer, Nucl.\ Phys.\ {\bf A526}, 703 (1991); 
A. R. Bodmer and
               C. E. Price, Nucl.\ Phys.\ {\bf A505}, 123 (1989).
%
\bibitem{Su94}   Y. Sugahara and H. Toki, 
Nucl.\ Phys.\ {\bf A579}, 557 (1994).
%
\bibitem{Re86} P. -G. Reinhard, M. Rufa, J. Maruhn, W. Greiner, and
                    J. Friedrich, Z. Phys.\  A {\bf 323}, 13 (1986).
%
\bibitem{Su95} K. Sumiyoshi, H. Kuwabara, and H. Toki,
               Nucl.\ Phys.\ {\bf A581}, 725 (1995).
%
\bibitem{Sh93} G. A. Lalazissis, J. K\"onig, and P. Ring,
               Phys.\ Rev.\ C {\bf 55}, 540 (1997).
%
\bibitem{Ga90} Y. K. Gambhir, P. Ring, and A. Thimet,
               Ann.\ Phys.\ (N.Y.) {\bf 198}, 132 (1990).
%
\bibitem{Ru88} M. Rufa, P. -G. Reinhard, J. A. Maruhn, W. Greiner, and
               M. R. Strayer, Phys.\ Rev.\ C {\bf 38}, 390 (1988).
%
\bibitem{Pa91} S. K. Patra and C. R. Praharaj, 
Phys.\ Rev.\ C {\bf 44}, 2552 (1991).
%
\bibitem{Va72} D. Vautherin and D. M. Brink, 
Phys.\ Rev.\ C {\bf 5}, 626 (1972).
%
\bibitem{De80} J. Decharg\'e and D. Gogny, 
Phys.\ Rev.\ C {\bf 21}, 1568 (1980).
%
\bibitem{Ho00} C. J. Horowitz and J. Piekarewicz,
               astro-ph/0010227 (2000).
%
\bibitem{Ko65} W. Kohn and L. J. Sham, 
 Phys.\ Rev.\ A {\bf 140}, 1133 (1965).
%
\bibitem{Sp92} C. Speicher, R. M. Dreizler, and E. Engel,
               Ann.\ Phys.\ (N.Y.) {\bf 213}, 413 (1992).
%
\bibitem{Br90} R. Brockmann and R. Machleidt, 
 Phys.\ Rev.\ C {\bf 42}, 1965 (1990).
%
\bibitem{serot86} C. J. Horowitz and B. D. Serot, 
                  Nucl.\ Phys.\ {\bf A368}, 503 (1981). 
%
\bibitem{boussy84} A. Bouyssy, S. Marcos, and Pham Van Thieu,
                  Nucl.\ Phys.\ {\bf A422}, 541 (1984).
%
\bibitem{bart82} J. Bartel, P. Quentin, M. Brack, C. Guet, and
                H. -B. H{\aa}kansson, 
                Nucl.\ Phys.\ {\bf A386}, 79 (1982).
%
\bibitem{Bay60}  G. Baym, Phys.\ Rev.\ {\bf 117}, 886 (1960);
                 B. M. Waldhauser, J. A. Maruhn, H. St\"ocker, and
                 W. Greiner, Phys.\ Rev.\ C {\bf 38}, 1003 (1988);
                 P. -G. Reinhard, Z. Phys.\  A {\bf 329}, 257 (1988).
%
\bibitem{negele70} J. W. Negele, Phys.\ Rev.\ C {\bf 1}, 1260 (1970).
%
\bibitem{li92} G. Q. Li, R. Machleidt, and R. Brockmann,
               Phys.\ Rev.\ C {\bf 45}, 2782 (1992).
%
\bibitem{poll99}L. Engvik, M. Hjorth-Jensen, R. Machleidt, H. M\"uther,
                and A. Polls,  Nucl.\ Phys.\ {\bf A627}, 85 (1997).
%
\bibitem{Ce93c} M. Centelles and X. Vi\~nas, 
Nucl.\ Phys.\ {\bf A563}, 173 (1993);
               M. Centelles, M. Del Estal, and X. Vi\~nas, 
               Nucl.\ Phys.\ {\bf A635}, 193 (1998).
%
%
\bibitem{egido96} T. Gonzalez-Llarena,  J. L. Egido, G. A. Lalazissis,
                  and  P. Ring,  Phys.\ Lett.\ B {\bf 379}, 13 (1996).
\bibitem{vretenar98} G. A. Lalazisis, D. Vretenar, and P. Ring,
                     Phys.\ Rev.\ C {\bf 57}, 2294 (1998).
\bibitem{farkan00} M. M. Sharma, A. R. Farhan, and S. Mythili,
                   Phys.\ Rev.\ C {\bf 61}, 054306 (2000).
%
\bibitem{Ch98} E. Chabanat, P. Bonche, P. Haensel, J. Meyer,
               and R. Schaeffer, Nucl.\ Phys.\ {\bf A635}, 231 (1998).
%
\bibitem{bohr1} A. Bohr and B. R. Mottelson, ``Nuclear Structure'',
              Vol.\ I (W.A. Benjamin, New York, 1969), Ch.\ 2.
%
\bibitem{tajima93} N. Tajima, P. Bonche, H. Flocard, P. -H. Heenen,
                   and M. S. Weiss, 
                   Nucl.\ Phys.\ {\bf A551}, 434 (1993).
%
\bibitem{sharma93} M. Sharma, G. A. Lalazissis, and P. Ring,
                   Phys.\ Lett.\ B {\bf 317}, 9 (1993).
%
%
%
%
%
%
\bibitem{doba96} J. Dobaczewski, T. R. Werner, J. F. Berger, 
                 C. R. Chenn, and
                 J. Deharge, Phys.\ Rev.\ C {\bf 53}, 2809 (1996); 
                 J. Terasaki, P. -H. Heenen, H. Flocard, and P. Bonche, 
                 Nucl.\ Phys.\ {\bf A600}, 371 (1996).
%
\bibitem{Page92} R. D. Page, P. J. Woods, R. A. Cunningham, 
                 T. Davidson, N. J. Davis, S. Hofmann, A. N. James, 
                 K. Livingston, P. J. Sellin, and A. C. Shotter, 
                 Phys.\ Rev.\ Lett.\ {\bf 68}, 1287 (1992). 
%
\end{thebibliography}
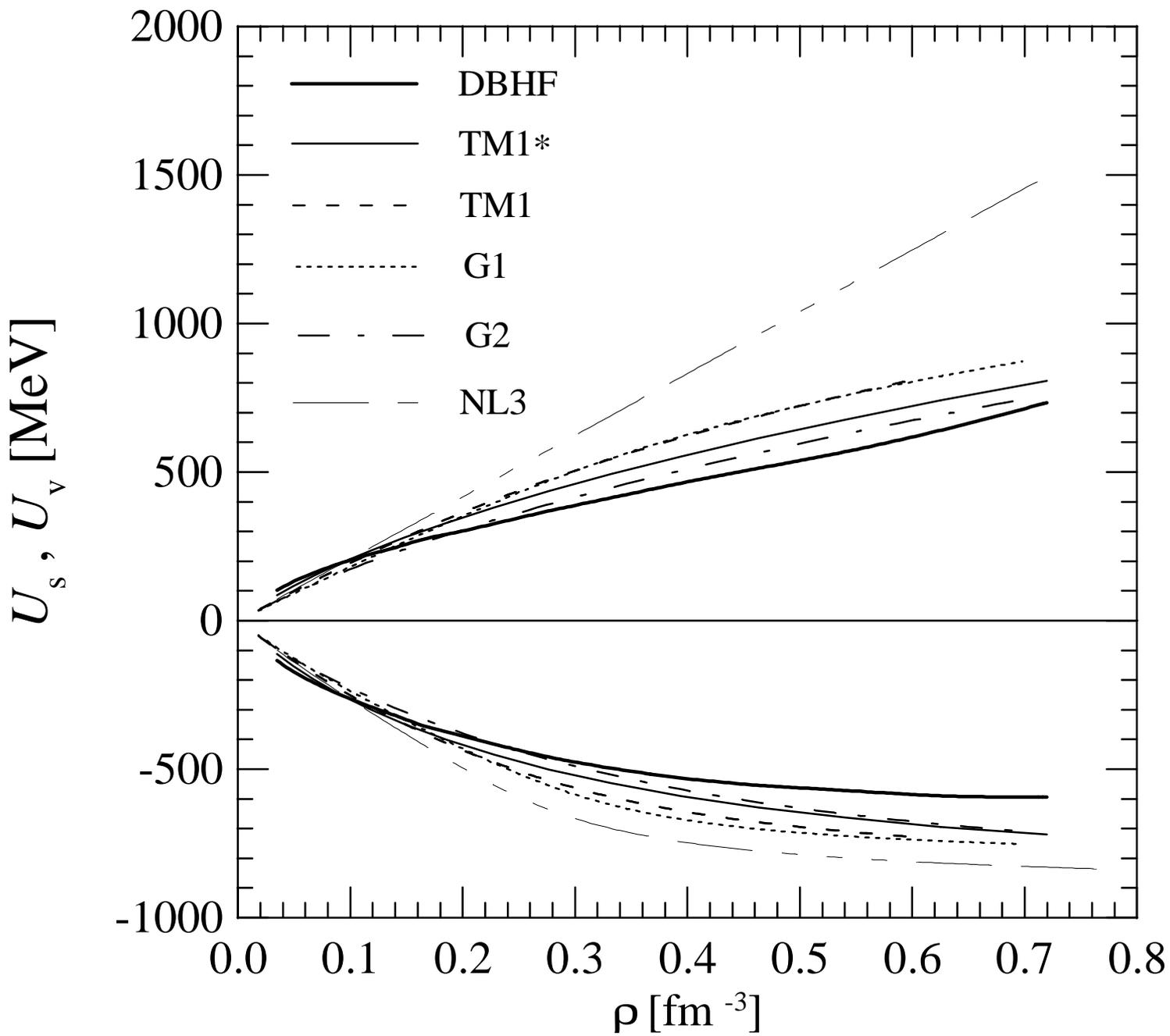
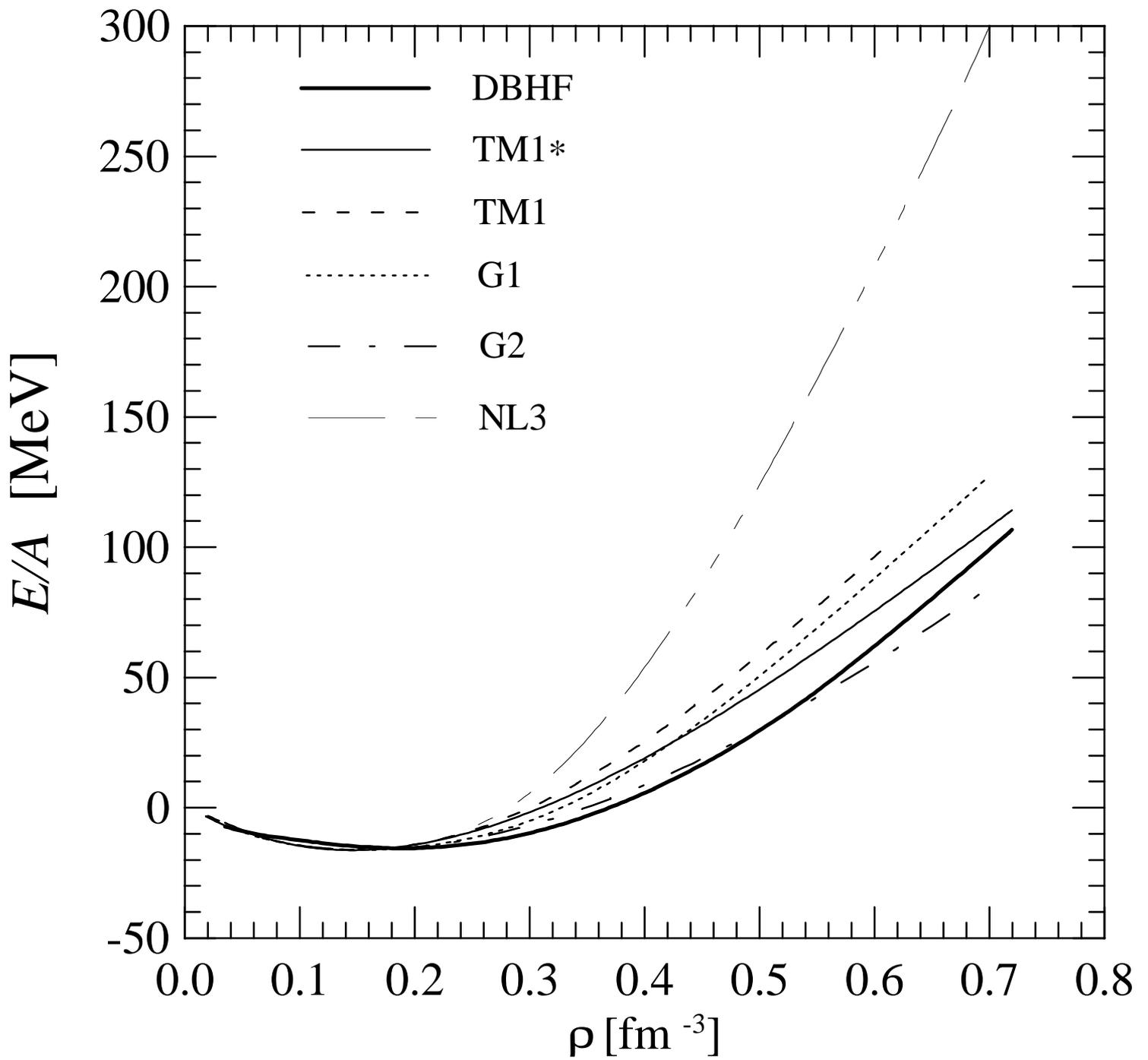
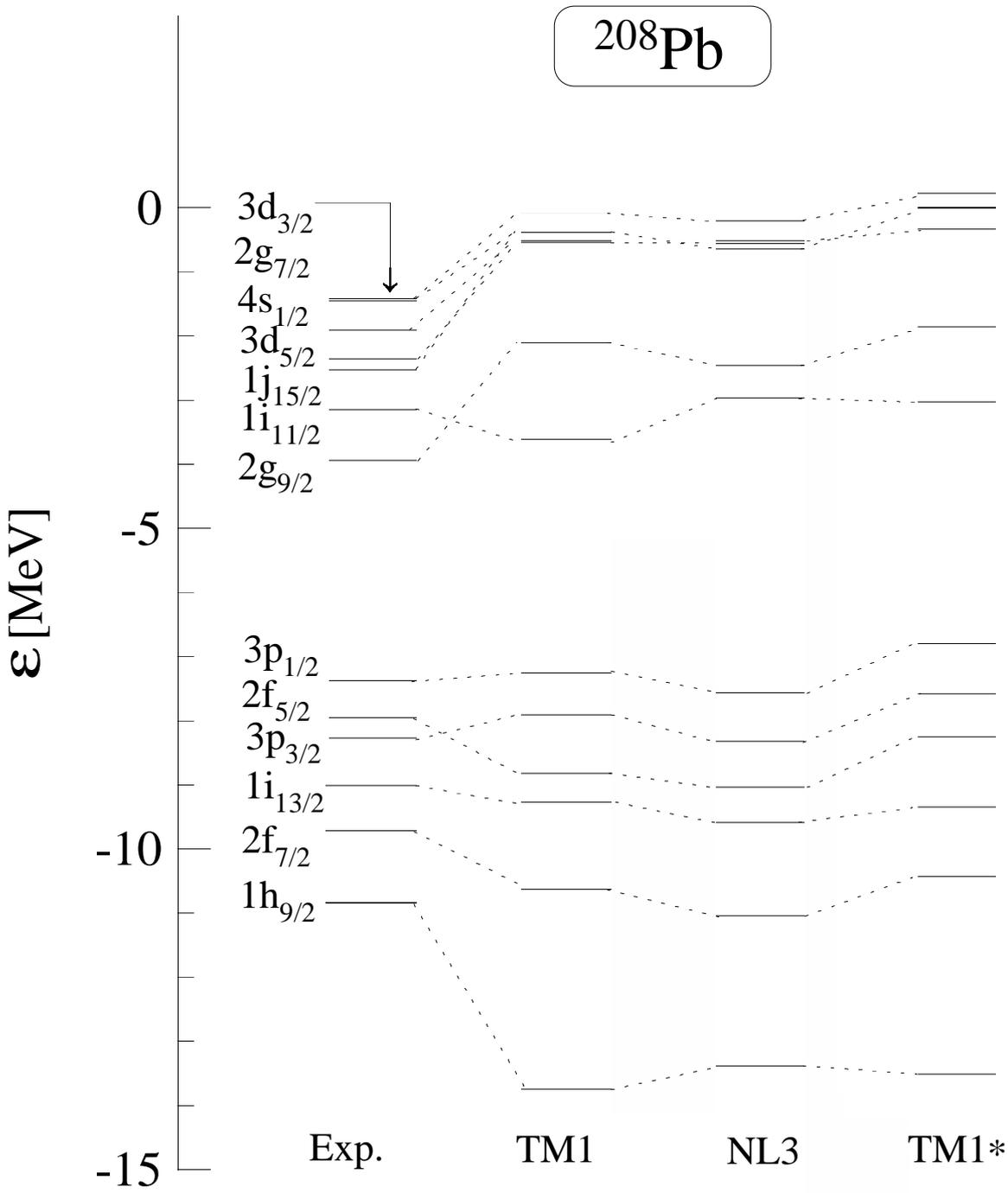
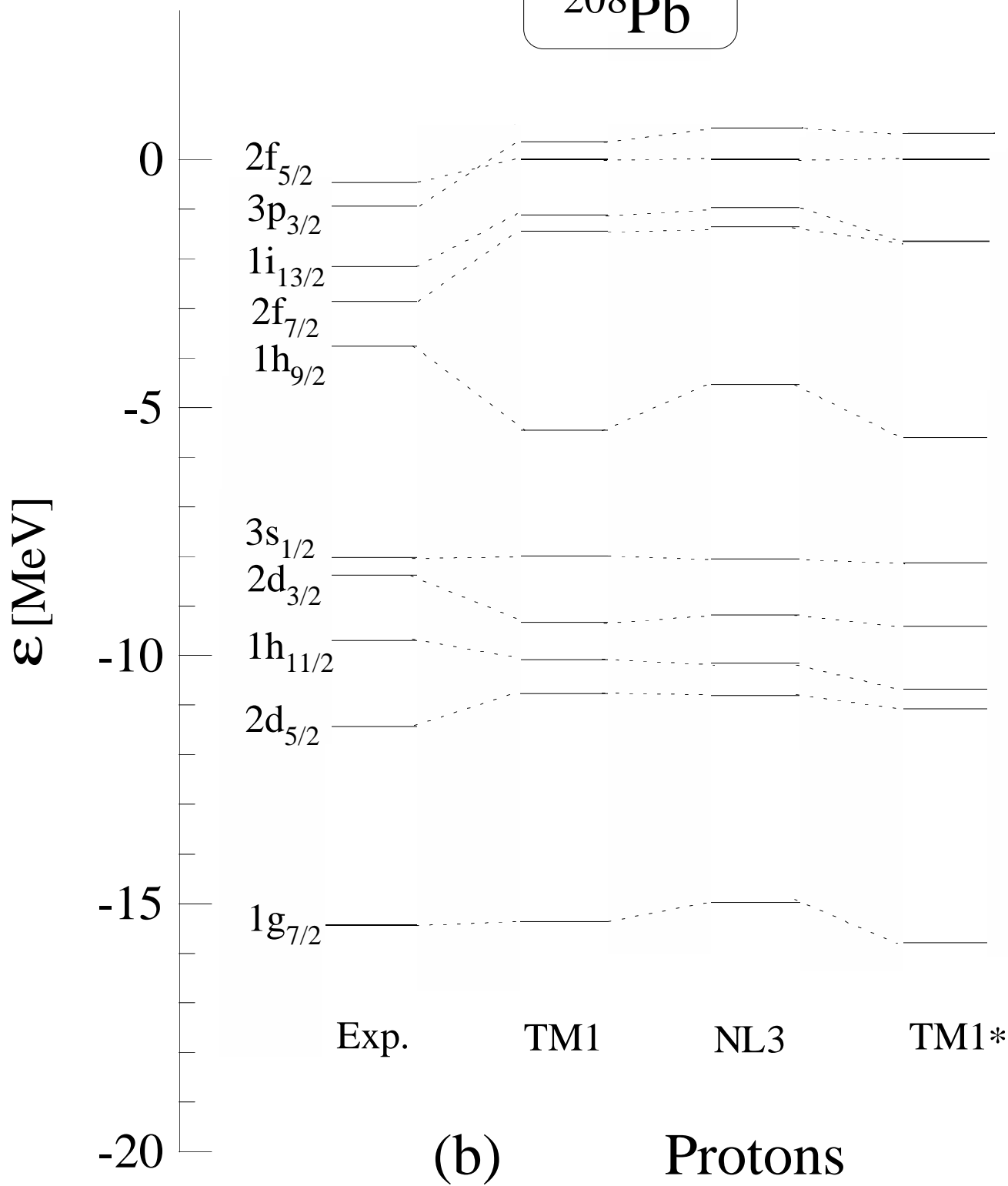
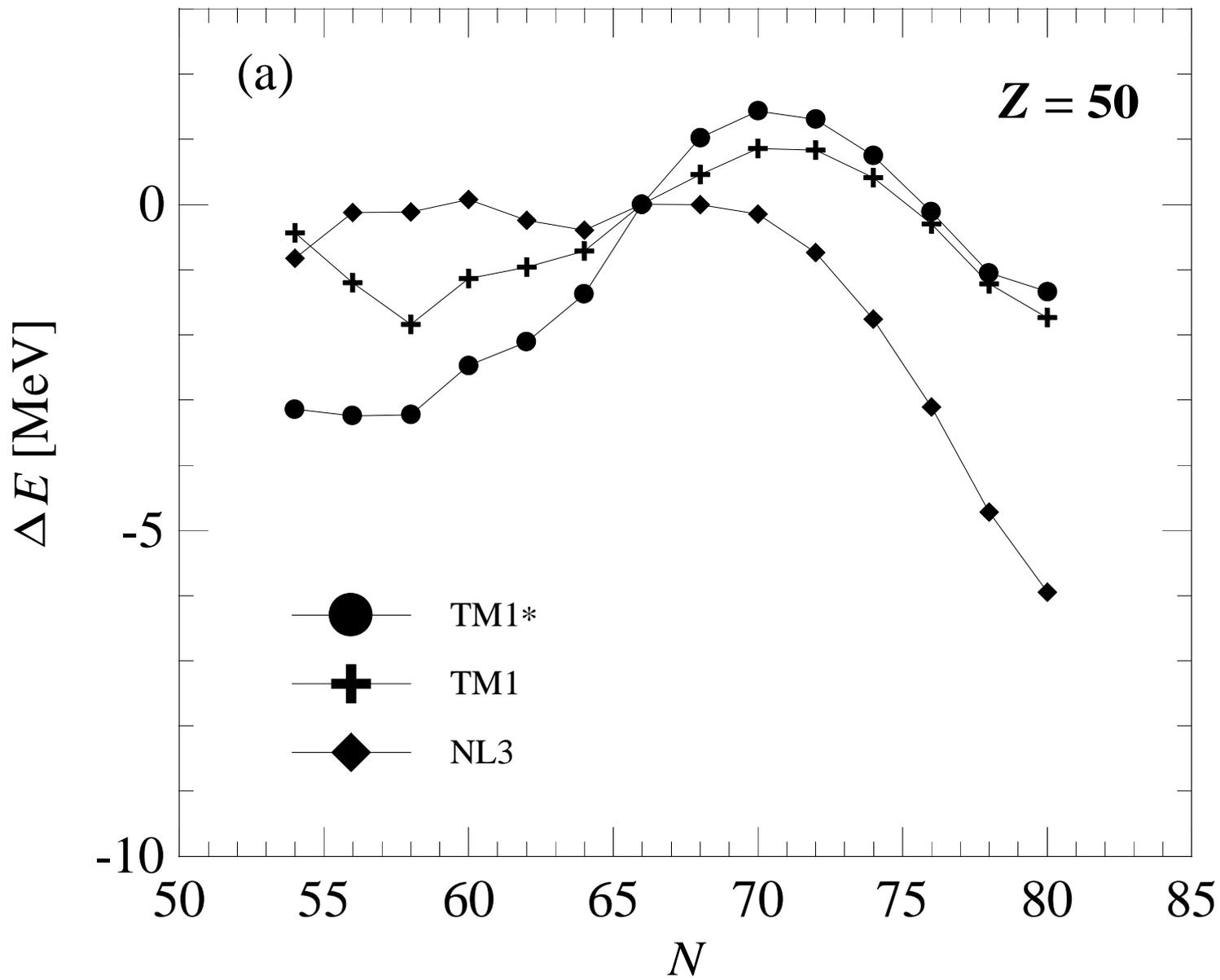
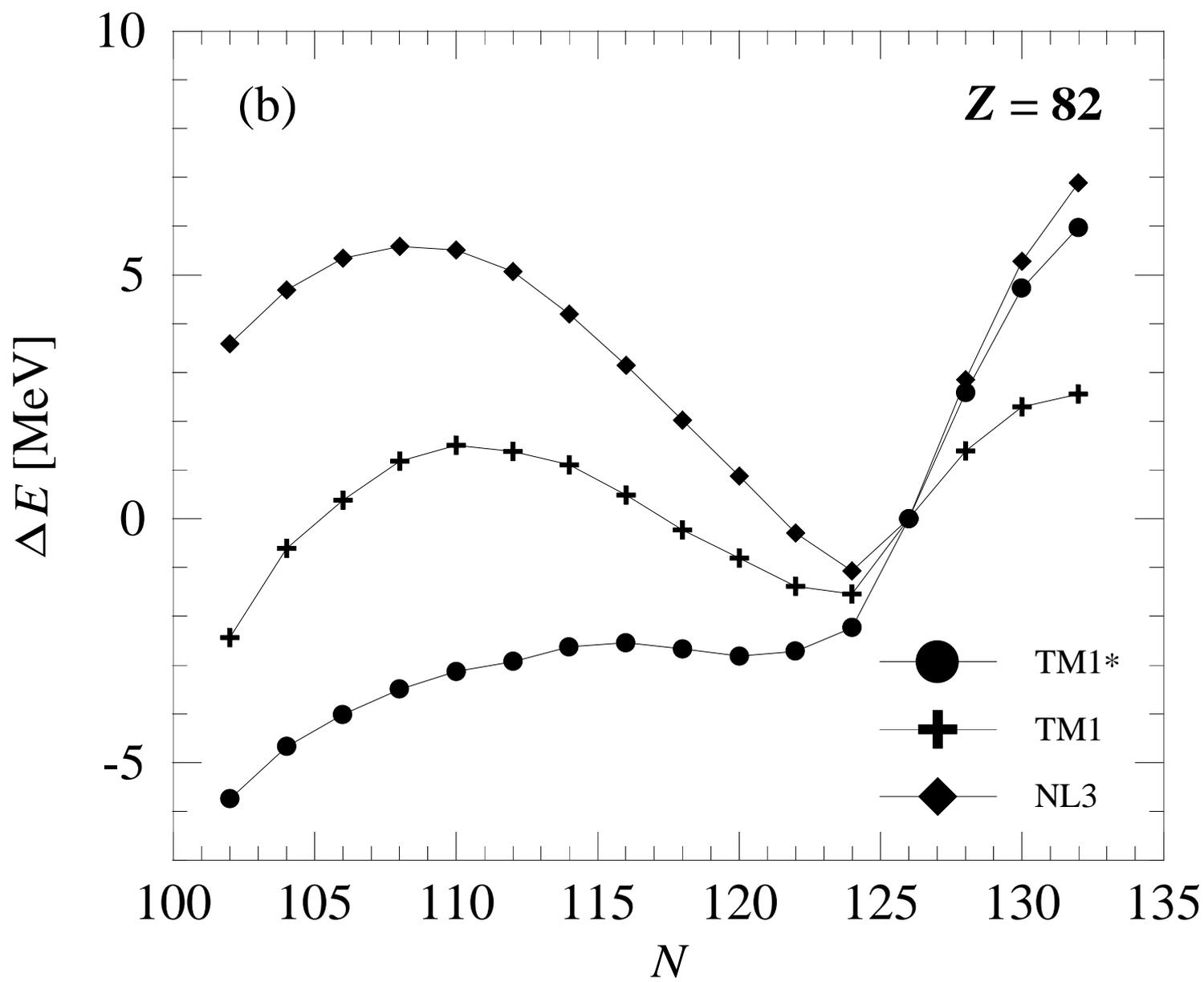
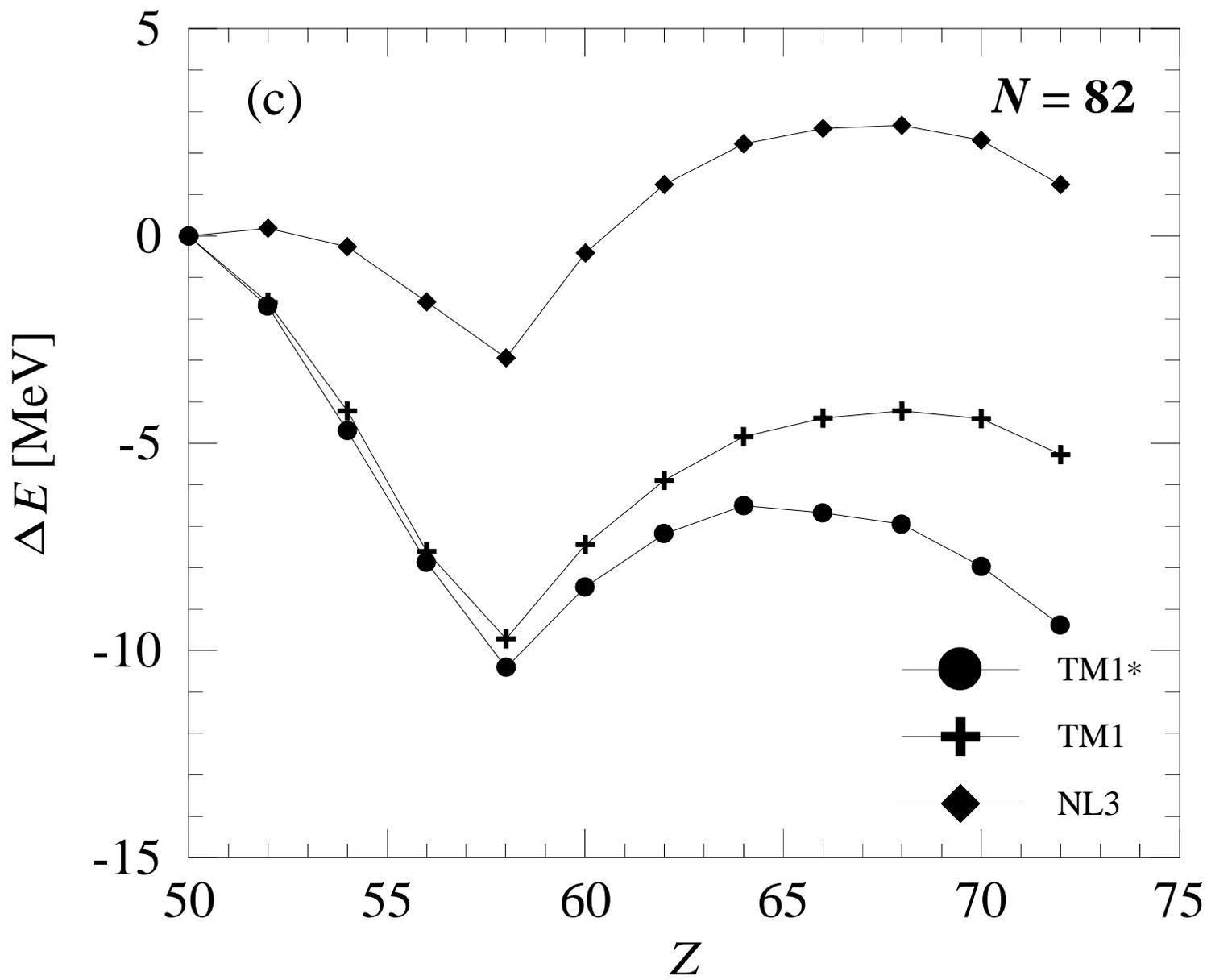
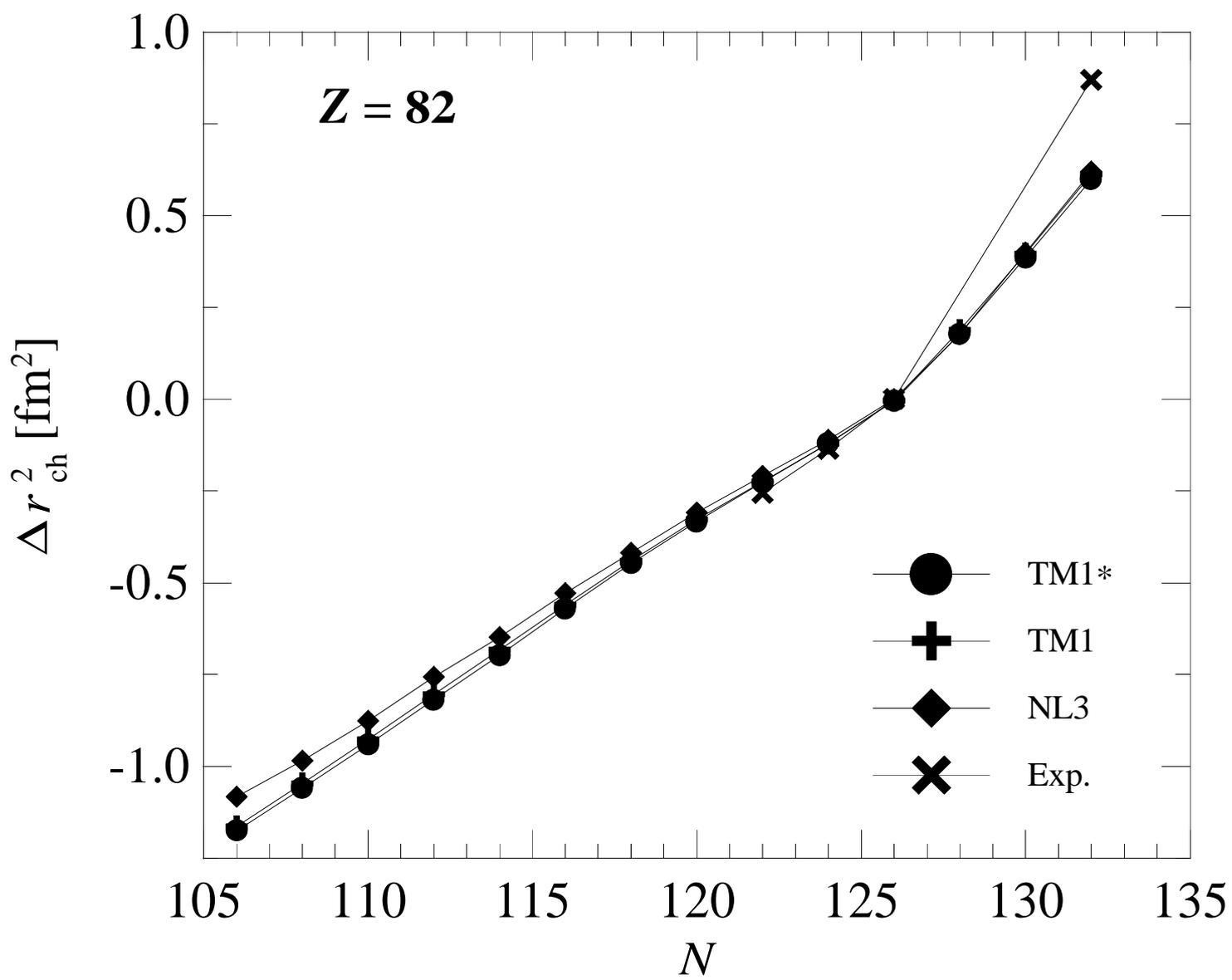
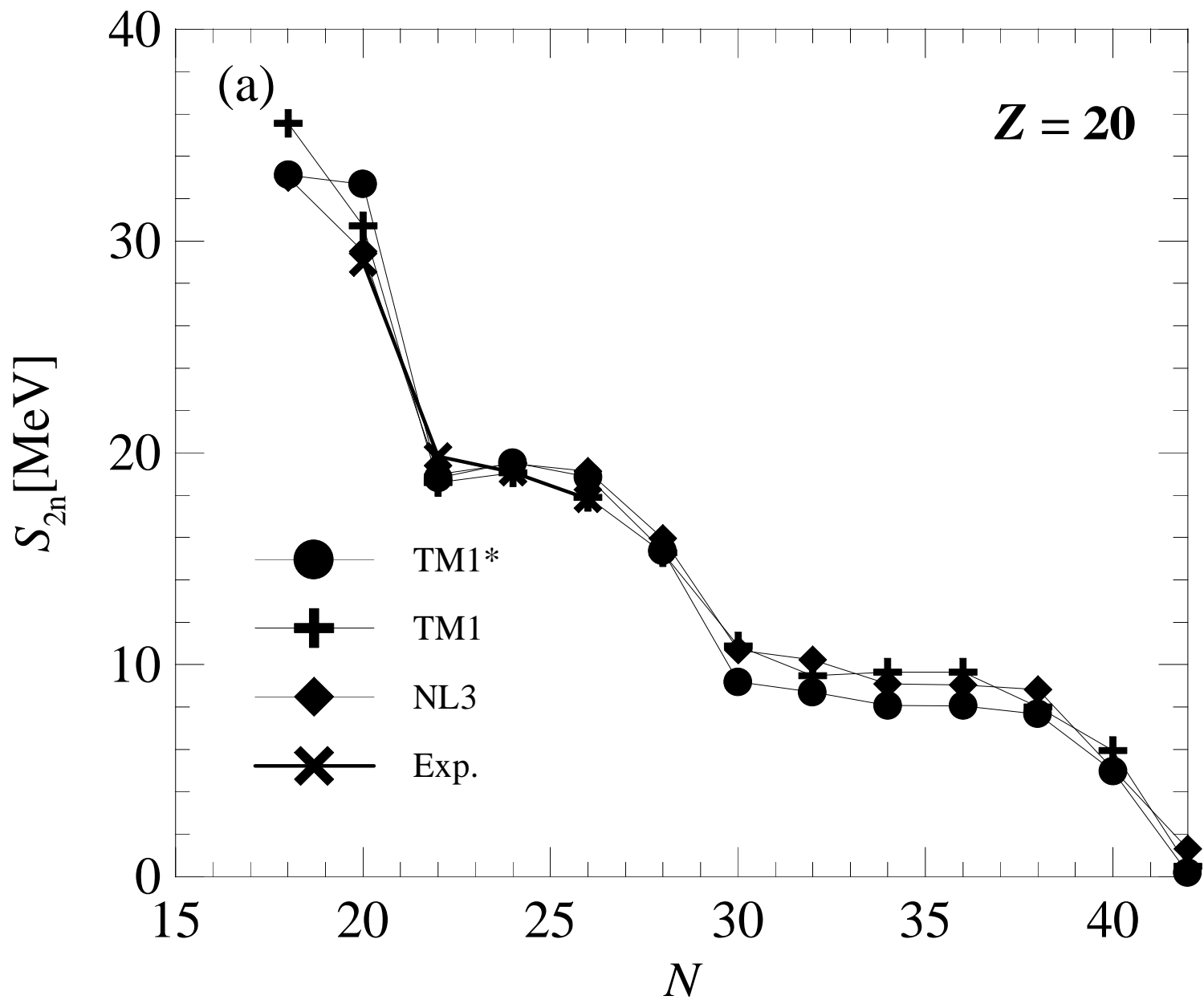
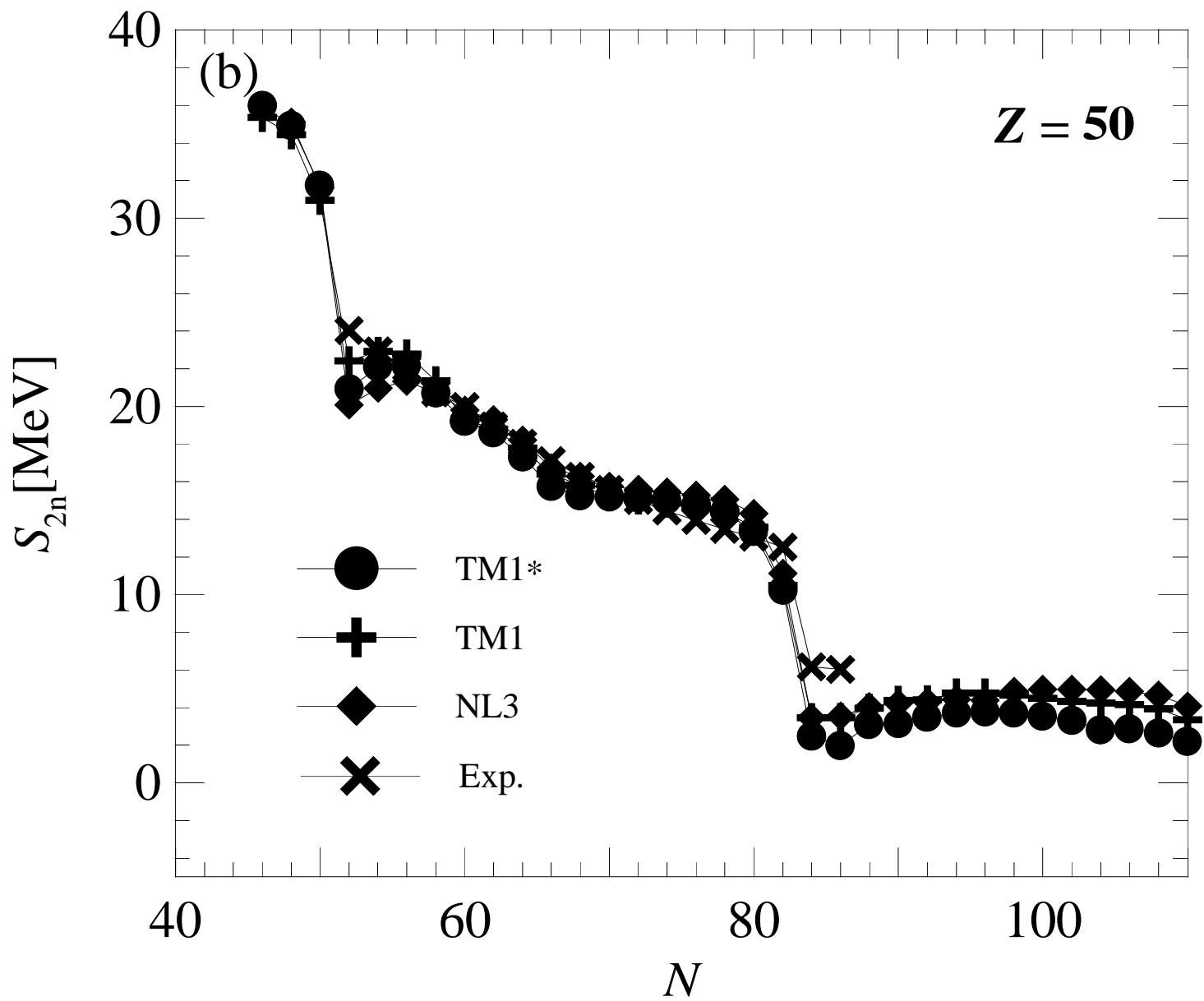
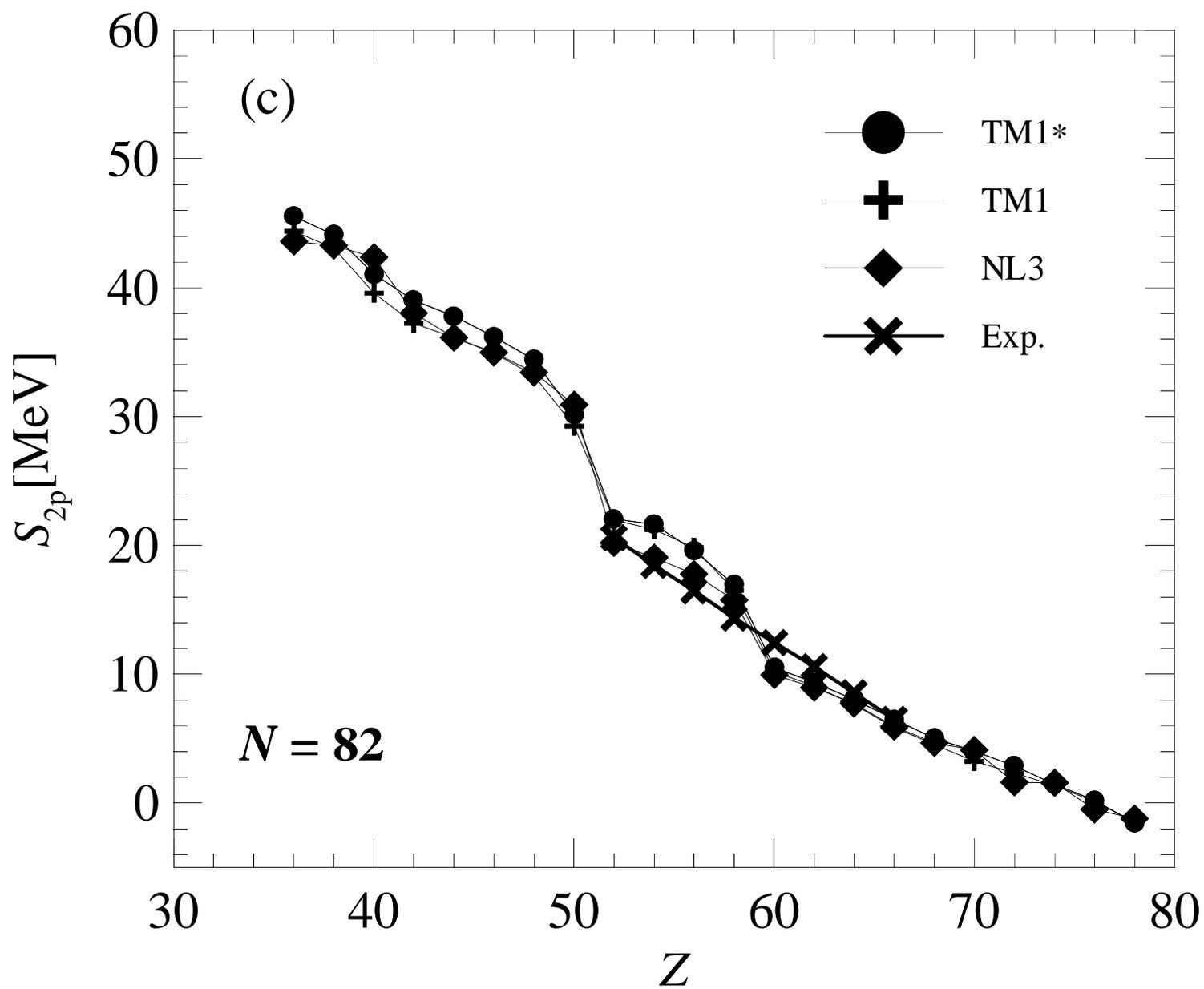
\end{document}